\documentclass[aps,preprint]{revtex4}
\usepackage{amsfonts}
\usepackage{amssymb,epsfig}
\usepackage{latexsym}
\usepackage{amsmath}
\usepackage{graphicx}

\begin{document}

\title{Agegraphic Dark Energy Model in Non-Flat Universe: Statefinder Diagnostic and $w-w^{\prime}$ Analysis.}
\author{M. Malekjani\footnote{%
E-mail: \texttt{malekjani@basu.ac.ir}}~~and\ A. Khodam-Mohammadi\footnote{%
E-mail: \texttt{khodam@basu.ac.ir}}}
\address{Physics Department, Faculty of Science, Bu-Ali Sina
University, Hamedan 65178, Iran}
\begin{abstract}
We study the interacting agegraphic dark energy (ADE) model in
non-flat universe by means of statefinder diagnostic and
$w-w^{\prime}$ analysis. First, the evolution of EoS parameter
($w_d$) and deceleration parameter ($q$) in terms of scale factor
for interacting ADE model in non-flat universe are calculated.
Dependence of $w_d$ on the ADE model parameters $n$ and $\alpha$ in
different spatial curvatures is investigated. We show that the
evolution of $q$ is dependent on the type of spatial curvature,
beside of dependence on parameters $n$ and $\alpha$. The accelerated
expansion takes place sooner in open universe and later in closed
universe compare with flat universe. Then, we plot the evolutionary
trajectories of the interacting ADE model for different values of
the parameters $n$ and $\alpha$ as well as for different
contributions of spatial curvature, in the statefinder parameters
plane. In addition to statefinder, we also investigate the ADE model
in non-flat universe with $w-w^{\prime}$ analysis.
\end{abstract}

\maketitle
%\pacs{04.20.Jb, 04.50.+h, 04.70.Bw, 04.70.Dy}

\section{Introduction\label{intro}}
In the framework of standard cosmology, the universe is dominated by
two mysterious component dark matter and dark energy. WMAP
(Wilkinson Microwave Anisotropy Probe) experiment \cite{wmap},
indicates that dark energy occupies about 70 \% of the total energy
of the universe, and the contribution of dark matter is $\sim$ 26\%.
The existence of dark matter is needed to explain the dynamics of
galaxies and the formation of structures in the universe
\cite{bos81}. While the dark energy component is responsible for
accelerating expansion of the universe. The dark energy (DE) problem
is one of the most famous problems in modern cosmology since the
discovery of accelerated expansion of the universe. The simplest
candidate for dark energy is the cosmological constant in which the
equation of state is independent of the cosmic time. This model is
the so-called $\Lambda$CDM, containing a mixture of cosmological
constant $\Lambda$ and cold dark matter (CDM). However, two problems
arise from this scenario, namely the fine-tuning and the cosmic
coincidence problems \cite{copel}. In order to solve these two
problems, many dynamical dark energy models were suggested, whose
equation of state evolves with cosmic time. The models with scale
fields such as quintessence \cite{wet88}, phantom field
\cite{cald02}, K-essence \cite{chi00} based on earlier work of K
-inflation \cite{arme99}, tachyon field \cite{sen02}, dilaton
\cite{gas02} and quintom \cite{eliza04} suggest that the energy with
negative pressure is provided by a scale field evolving down a
proper potential. Also the interacting dark energy models including
Chaplygin gas \cite{kame01}, holographic dark enrgy models
\cite{coh99}, and braneworld models \cite{def02} have been proposed.
Recently, based on principle of quantum gravity, the
 agegraphic dark energy (ADE) and the new agegraphic dark
energy (NADE) models were proposed by Cai \cite{Cai1} and Wei \& Cai
\cite{cai4}, respectively. The ADE model is based on the line of
quantum fluctuations of spacetime, the so-called K\'{a}rolyh\'{a}zy
relation $\delta t=\lambda t_{p}^{2/3}t^{1/3}$, and the energy-time
Heisenberg uncertainty relation $E_{\delta t^{3}}\sim t^{-1}$. These
relations enable one to obtain an energy density of the metric
quantum fluctuations of Minkowski spacetime as follows \cite{Maz}
\begin{equation}
\rho _{q}\sim \frac{E_{\delta t^{3}}}{\delta t^{3}}\sim \frac{1}{%
t_{p}^{2}t^{2}}\sim \frac{m_{p}^{2}}{t^{2}}.  \label{ED}
\end{equation}%
 In ADE model the energy density of dark energy is given by Eq.(\ref{ED}).
 However, in Friedmann-Robertson-Walker (FRW) universe, due to effect of
curvature, one should assume a numerical factor $3n^{2}$ in
Eq.(\ref{ED}) \cite{Cai1}. The new model of agegraphic dark energy
(NADE) has been proposed by Wei and Cai \cite{cai4}, in which the
cosmic time is replaced by the conformal time. The recent
observational data from the Abell Cluster $A586$ support the
interaction between dark matter and dark energy \cite{bertol}.
However, the strength of this interaction is not exactly identified
\cite{feng}. The ADE and NADE models have been extended regarding
the interaction between dark components of the universe, in Refs.
\cite{wei09,kim08,shaikh,wei08}. Also, the observational experiments
such as CMB experiments \cite{Sie}, and the luminosity-distance of
supernova measurements \cite{Caldwell} reveal the non-flat universe
with tiny positive curvature. Hence, we are motivated to consider
the
non-flat universe containing the interacting dark matter and dark energy components.\\
As was mentioned above, the various dark energy models have been
proposed to describe the accelerated phase of the universe. The
property of dark energy in these models is strongly model dependent.
In order to be capable of differentiating between various models, a
sensitive diagnostic tool is needed. The geometrical statefinder
diagnostic tool that makes use of parameter pair \{$s,r$\},
introduced by Sahni et al. \cite{Sahni}, can discriminate various
dark energy models. The statefinder pair is constructed directly
from a spacetime metric. The importance of such pair is to
distinguish of the cosmological evolution behaviors of dark energy
models with the same values of $H_{0}$ and $q_{0}$ at the present
time. At future by combining the data of Supernova acceleration
probe (SNAP) with statefinder diagnosis, we may choose the best
model of dark energy. The statefinder pair \{$s,r$\} are given by
\cite{Sahni}
\begin{equation}
r=\frac{\dddot{a}}{aH^{3}},~~~~~~~s=\frac{r-1}{3(q-1/2)},
\label{rspair}
\end{equation}

 Up to now, many authors have studied the dark energy models by means of
 statefinder diagnostic analysis. The standard $%
\Lambda $CDM model and quintessence \cite{Sahni,r10}, interacting
quintessence models \cite{r12,r13}, the holographic dark energy
models
\cite{r14,r15}, the holographic dark energy model in non-flat universe \cite%
{r16}, the phantom model \cite{r18}, the tachyon \cite{r22}, the ADE
model with and without interaction in flat universe \cite{wei} and
the interacting NADE model in flat and non-flat universe
\cite{zhang,khod10}, are analyzed through the statefinder
diagnostic.
 In addition to the statefinder diagnostic, another analysis to
 discriminate various models of dark energy is $w-w^{\prime }$ analysis
which is used widely in the literature \cite{wei,wwp}.\\
In this paper we study the ADE model in a non-flat universe. First,
the  evolutionary  behavior of EoS parameter, $w_d$, and
deceleration parameter $q$ for different illustrative values of
model parameters $n$ and $\alpha$ ($\alpha$ is the interaction
parameter between dark matter and dark energy) in non-flat universe
are investigated. Then we examine the ADE model in a non-flat
universe by means of statefinder diagnostic tool and $w-w^{\prime}$
analysis.

\section{The ADE Model in Non-Flat Universe \label%
{theory}}

As we mentioned in Sec. (\ref{intro}), the energy density of dark
energy in ADE model is given by
\begin{equation}
\rho _{d}=\frac{3n^{2}m_{p}^{2}}{T^{2}}  \label{adedens}
\end{equation}%
where the cosmic time $T$ is defined as the age of the universe%
\begin{equation}
T=\int_0^t dt=\int_{0}^{a}\frac{da}{Ha}.
\end{equation}%

Assuming a non-flat Friedman-Robertson-Walker (FRW) universe
containing the agegraphic dark energy and cold dark matter
components, the corresponding freidmann equation is as follows
\begin{equation}
H^{2}+\frac{k}{a^{2}}=\frac{1}{3m_{p}^{2}}(\rho _{m}+\rho _{d})  \label{Freq}
\end{equation}%
where $k=1,0,-1$ is curvature parameter corresponding to a closed,
flat and open universe, respectively. Recent observations support a
closed universe with a tiny positive small curvature $\Omega
_{k0}=1/H_{0}^{2}\simeq 0.02$ \cite{wmap}. The other form of
Friedmann equation with respect to fractional energy density $\Omega
_{i}=\rho _{i}/\rho _{c}$, and critical density $\rho
_{c}=3m_{p}^{2}H^{2}$ is
\begin{equation}
\Omega _{m}+\Omega _{d}=1+\Omega _{k}.  \label{Freq2}
\end{equation}%
From Eq.(\ref{adedens}), it is easy to find that
\begin{equation}
\Omega_d=\frac{n^2}{H^2T^2} \label{omeg_d}
\end{equation}
The continuity equations for interacting dark energy and dark matter
are given by
\begin{eqnarray}
\dot{\rho _{m}}+3H\rho _{m} &=&Q, \\
\dot{\rho _{d}}+3H(\rho _{d}+p_{d}) &=&-Q,  \label{contd}
\end{eqnarray}%
 Three forms of $Q$ which have been extensively used in the literatures \cite{cai4, zhang, shikh} are%
\begin{equation}
Q=9\alpha _{i}m_{p}^{2}H^{3}\Omega _{i};\qquad \Omega _{i}=\left\{
\begin{array}{ll}
\Omega _{d}; & i=1 \\
\vspace{0.75mm}\Omega _{m}; & i=2 \\
\Omega _{d}+\Omega _{m}; & i=3%
\end{array}%
\right. .  \label{qform}
\end{equation}%
Differentiating Eq.(\ref{omeg_d}) and using Eqs.(\ref{adedens}), (\ref%
{Freq}), (\ref{Freq2}) and (\ref{contd}), the derivative of $\Omega
_{d}$ can be calculated
as%
\begin{eqnarray}
\Omega _{d}^{\prime } &=&\frac{\dot{\Omega}}{H}=-2\Omega _{d}\left[ \frac{%
\dot{H}}{H^{2}}+\frac{\sqrt{\Omega _{d}}}{n}\right] ;  \label{OmP} \\
\frac{\dot{H}}{H^{2}} &=&-\frac{\Omega
_{d}^{3/2}}{n}-\frac{3}{2}(1-\Omega _{d})-\frac{\Omega
_{k}}{2}+\frac{Q_{c}}{2},  \label{HD}
\end{eqnarray}%
where prime denotes the derivative with respect to $\ln a$ and $%
Q_{c}=Q/H\rho _{c}=3\alpha _{i}\Omega _{i}.$ The relations (\ref{OmP}) and (%
\ref{HD}), also has been obtained in \cite{shaikh} for third
interaction form of $Q.$

Substituting the relation (\ref{HD}) in Eq.(\ref{OmP}), we obtain a
normal differential equation for $\Omega _{d}$ as%
\begin{equation}
\Omega _{d}^{\prime }=\Omega _{d}\left[ (1-\Omega _{d})(3-2\frac{\sqrt{%
\Omega _{d}}}{n})+\Omega _{k}-Q_{c}\right] ,  \label{Oddiff}
\end{equation}%
where $\Omega _{k}$ is given by
\begin{equation}
\Omega _{k}=\frac{a\gamma (1-\Omega _{d})}{1-a\gamma }.  \label{Ok}
\end{equation}%
Here $\gamma $ satisfies the following equation%
\begin{equation}
\frac{\Omega _{k}}{\Omega _{m}}=a\frac{\Omega _{k0}}{\Omega _{m0}}=a\gamma .
\end{equation}%
From the continuity equation (i.e., Eq.\ref{contd}) and
Eqs.(\ref{adedens},\ref{omeg_d}), it is easy to find that the EoS
parameter of the interacting agegraphic dark energy,
$w_d=p_d/\rho_d$, can be obtained  as
\begin{equation}
w_{d}=-1-Q_{cd}+\frac{2\sqrt{\Omega _{d}}}{3n},  \label{EoS}
\end{equation}%
where $Q_{cd}=Q_{c}/(3\Omega _{d})$.  Using the Eq.(\ref{Oddiff}),
the evolutionary behavior of $w_{d}$ is given by the following
equation
\begin{equation}
w_{d}^{\prime }=+\frac{\sqrt{\Omega _{d}}}{3n}\left[ 1+\Omega
_{k}-3\Omega _{d}-2\frac{\sqrt{\Omega _{d}}}{n}(1-\Omega
_{d})-Q_{c}\right] -Q_{cd}^{\prime }.  \label{wp}
\end{equation}%

The deceleration parameter $q$ can be obtained as
\begin{eqnarray}
q &=&-\frac{\dot{H}}{H^{2}}-1=\frac{(1+\Omega _{k})}{2}+\frac{3}{2}\Omega
_{d}w_{d}  \label{q} \\
&=&\frac{\Omega _{d}^{3/2}}{n}-\Omega _{d}+\frac{1-\Omega
_{d}}{2(1-a\gamma )}-\frac{Q_{c}}{2}  \label{q2}
\end{eqnarray}%
$\allowbreak $

\section{STATEFINDER DIAGNOSTIC FOR INTERACTING ADE MODEL IN  NON-FLAT UNIVERSE
\label{STF}}

Now we switch to the statefinder pair \{r,s\}, which was expressed in Sec.\ref%
{intro}. From the definition of $q$ and $H$, the parameter $r$ in Eq.(\ref{rspair}%
) can be written as
\begin{equation}
r=\frac{\ddot{H}}{H^{3}}-3q-2.  \label{r}
\end{equation}%
Using Eqs.(\ref{HD}), (\ref{EoS}) and (\ref{q}), we have%
\begin{equation}
\frac{\ddot{H}}{H^{3}}=\frac{9}{2}+\frac{9}{2}\Omega
_{d}w_{d}(w_{d}+Q_{cd}+2)-\frac{3}{2}\Omega _{d}w_{d}^{\prime }+\frac{5}{2}%
\Omega _{k}.  \label{Hdd}
\end{equation}%
Hence, the Eq.(\ref{r}) can be obtained as%
\begin{equation}
r=1+\Omega _{k}+\frac{9}{2}\Omega _{d}w_{d}(w_{d}+Q_{cd}+1)-\frac{3}{2}%
\Omega _{d}w_{d}^{\prime }.  \label{r2}
\end{equation}
In a non flat universe, Evans et al. \cite{Evans} generalize the
definition of parameter $s$ in Eq.(\ref{rspair}) as
\begin{equation}
s=\frac{r-\Omega _{tot}}{\frac{3}{2}(q-\frac{\Omega _{tot}}{2})},  \label{sg}
\end{equation}%
where the total fractional energy density is $\Omega _{tot}=\Omega
_{m}+\Omega
_{d}=1+\Omega _{k}$.  Therefore from this new definition we have%
\begin{equation}
s=1+w_{d}+Q_{cd}-\frac{w_{d}^{ \prime }}{3w_{d}}.  \label{sk}
\end{equation}
By omitting $w_{d}^{\prime }$ between (\ref{r2}) and (\ref{sk}), we
find $r$ in terms of $s $ as follows
\begin{equation}
r=1+\Omega _{k}+\frac{9}{2}s\Omega _{d}w_{d}.  \label{r3}
\end{equation}

\section{Numerical results\label{NR}}
In this section, first we calculate the EoS parameter, $w_d$, of ADE
model and deceleration parameter, $q$, for different model
parameters $n$ and $\alpha$ in non flat universe. Then we study the
ADE model in non flat universe by means of statefinder diagnostic
tool and $w-w^{\prime}$ analysis. Here we assume the first case of
interaction form, $Q=Q_{1}=9\alpha m^{2}H^{3}\Omega _{d}$, in
Eq.(\ref{qform}). In this case, $Q_c=3\alpha \Omega_d$,
$Q_{cd}=\alpha$ and the differential equation for $\Omega _{d} $
(i.e., Eq.\ref{Oddiff}) can be reduced as
\begin{equation}
\Omega _{d}^{\prime }=\Omega _{d}[(1-\Omega _{d})(3-\frac{2\sqrt{\Omega _{d}}%
}{n})-3\Omega _{d}\alpha +\Omega _{k}]  \label{Omeg_d}
\end{equation}

The Eqs.(\ref{EoS}), (\ref{wp}), (\ref{r2}) and (\ref{sk}%
), reduced as follows
\begin{equation}
w_{d}=-1+\frac{2\sqrt{\Omega _{d}}}{3n}-\alpha   \label{w_d}
\end{equation}
\begin{equation}
w_{d}^{\prime }=\frac{\sqrt{\Omega _{d}}}{3n}[(1+\Omega
_{k})-3\Omega _{d}(1+\alpha )-\frac{2\sqrt{\Omega _{d}}}{n}(1-\Omega
_{d})]  \label{wp_d}
\end{equation}
\begin{equation}
r=1+\Omega _{k}+\frac{9}{2}\Omega _{d}w_{d}(w_{d}+\alpha +1)-\frac{3}{2}%
\Omega _{d}w_{d}^{\prime }.  \label{state_r}
\end{equation}
\begin{equation}
s=1+w_{d}+\alpha -\frac{w_{d}^{\prime }}{3w_{d}}.  \label{state_s}
\end{equation}

From Eq.(\ref{w_d}), it is easy to see that in the absence of
interaction ($\alpha=0.0$) $w_d$ is larger than $-1$, and the ADE
model can not cross the phantom divide. In the presence of
interaction between dark energy and dark matter the ADE model can
cross the phantom divide, if $\alpha>2\sqrt{\Omega_d}/3n$. Taking
$\Omega_d=0.72$ for the present time, the phantom-like EoS can be
obtained if $n\alpha> 0.565$. The WMAP and SDSS observational
experiments indicate that the best value for $n$ is $3.4$
\cite{amend99}. Thus, the condition $w_d<-1$ leads to
$\alpha>0.166$. For example for $\alpha=0.2$ we get $w_d=-1.03$ at
the present time. Hence, the phantom-like equation of state can be
generated from an
interacting ADE model in the universe with any spacial curvature.\\

Fig.(1) shows the evolution of $w_{d}$ in terms of scale factor,
$a$, for different model parameters $n$ and $\alpha$ and various
spatial curvatures. In first arrow panels, $w_d$ is plotted in the
absence of interaction term between dark matter and dark energy
($\alpha=0.0$). In this case the phantom divide can not be crossed
for any spatial curvature. In second and third arrow panels, the
interaction between dark matter and dark energy is taking into
account. For $\alpha=0.1$, the phantom divide is achieved for
various spatial curvature. Increasing the parameter $n$ leads to
smaller value of $w_d$. In third arrow, $\alpha=0.2$, the treatment
of $w_d$ may be different form previous cases. In the case of
($n=0.9, \alpha=0.2$), $w_d$ is positive at the early time, $w_d>0$.
This behavior of $w_d$ at the early time is due to the presence of
$2\sqrt{\Omega_d}/3n$ in Eq.(\ref{w_d}). This term leads to larger
value of $w_d$ for smaller value of $n$. It should be noted that for
smaller value of $n$ and larger value of $\alpha$, we get a larger
value of $\Omega_d$ at the early time (see Fig.(1) of
Ref.\cite{wei09}). Hence, it is easy to see that $w_d$ tends to the
larger values because of the presence of $n$ at denominator and
$\Omega_d$ at numerator of Eq.(\ref{w_d}). For instance for the
model parameters $n=0.9$,$\alpha=0.2$, at the scale factor $a=0.01$,
$2\sqrt{\Omega_d}/3n=1.3$  which is bigger than $1+\alpha=1.2$, thus
$w_d>0$. For the model parameters $n=1.2$, $\alpha=0.2$, the ADE
model can cross the phantom divide in the flat and closed universe,
but in open universe it can not cross the phantom divide. In the
case of ($n=2$, $\alpha=0.2$) the phantom divide can be achieved for
any spacial curvature. By comparing with NADE model which has been
investigated by us \cite{khod10}, we see that in ADE model, for a
given value of $\alpha$, the phantom divide  is achieved for smaller
$n$ at early time. For example, for $\alpha=0.1$, the ADE model with
$n=0.9$ can cross the phantom divide at the early time (see Fig.1),
while in NADE model it has been achieved, if $n>3$ (see Fig.1-b of Ref.\cite{khod10}).\\

The other cosmological parameter which we calculate it, is the
deceleration parameter $q$. The parameter $q$ in ADE model is given
by Eq.(\ref{q}). In the early time, where $\Omega _{d}\rightarrow 0$
and $\Omega _{k}\rightarrow 0$, the parameter $q$ converges to
$1/2$, whereas the universe has been dominated by dark matter. In
Fig(2), we show the evolution of $q$ as a function of cosmic scale
factor for different model parameters $n$ and $\alpha$ and also
various contribution of spatial curvature of the universe. In first
arrow panels, the ADE model without interaction term ($\alpha=0.0$),
we see that increasing the parameter $n$ leads to smaller value for
$q$. The transition from decelerated expansion ($q>0$) to
accelerated expansion ($q<0$) tacks place sooner for larger value of
$n$. For the parameters $n=0.9$, $\alpha=0.0$,  the accelerated
universe can not be archived even at the late time. The transition
from decelerated to the accelerated universe occurs gradually from
open, flat and closed universe. However, the difference between them
is very small, but we can interpret that the accelerated phase
occurs earlier in open universe. In second arrow panels, we consider
$\alpha=0.1$. Here, in this case, the universe  enters into the
accelerated phase earlier compare with the pervious case
($\alpha=0.0$). In third arrow panels, the interaction parameter is
assumed as $\alpha=0.2$. We see that in this case the accelerated
phase is achieved sooner than previous cases. Hence, the universe
enters the accelerated phase earlier by increasing the interaction
parameter $\alpha$.\\

At following, we calculate the evolutionary trajectories in the
statefinder plane and analyze the interacting ADE model in non-flat
universe with statefinder point of view. The statefinder is a
geometrical diagnostic tool, because it depends only on the scale
factor $a$. The standard $\Lambda$CDM model corresponds to a fixed
point \{r=1, s=0\} in the r-s diagram  in a flat universe
\cite{Sahni}, and \{$r=1+\Omega _{k}$,$s=0$\} in a non-flat universe
\cite{Evans}. It should be mentioned that the statefinder diagnostic
for ADE model in flat universe has been investigated in
Ref.\cite{wei}, where the focus is put on the diagnostic of the
different values of parameters $n$ and $\alpha$. In Ref.\cite{wei},
it has been discussed that from the statefinder viewpoint $n$ and
$\alpha$ play the significant role in this model and it leads to the
values of \{r, s\} in today and future tremendously different. Here
we want to focus on the statefinder
diagnostic of the spatial curvature contribution in the ADE model.\\
The Eqs.(\ref {state_r},\ref{state_s}) and (\ref{wp_d}) describe the
evolution of statefinder parameters \{$r$,$s$\} and also
$w_{d}^{\prime }$. Since $\Omega_d\rightarrow 0$ at the early time,
thus from Eq.(\ref{wp_d}) one can see that $ w_{d}^{\prime
}\rightarrow 0,$ at this time. It is worth noting that $\Omega_k$
tends to zero at the early time. Using Eqs. (\ref
{state_r},\ref{state_s}), we see that the ADE model in the early
time, independent of model parameters $n$ and $\alpha$ and any
contribution of spatial
curvature, gives the fixed point ($r=1$,$s=0$) in $r-s$ diagram.\\
In Fig.(3) the evolutionary trajectories of statefinder for the
interacting ADE model is plotted. While the universe expands, the
trajectories  of the statefinder start from the fixed point \{r=1,
s=0\} at the early time and then the parameter $s$ increases and the
parameter $r$ decreases. In first arrow panels, the ADE is
considered without interaction term ($\alpha=0.0$). We see that, in
addition to the model parameters $n$ and $\alpha$, the evolutionary
trajectory is dependent on the model curvature of the universe. The
colore points on the curves represent the today's values of
statefinder parameters ($r_0,s_0$). The present value $r_0$ in
closed universe is largest, while $s_0$ is smallest compare with
flat and open universe. Also, larger value of $n$ obtains the larger
$r_0$ and smaller $s_0$. In second arrow panels, the interaction
parameter is chosen as $\alpha=0.1$. In this case, we see that the
present values ($r_0$,$s_0$) are smaller compare with the case of
$\alpha=0.0$. In third arrow panels the interaction parameter is
$\alpha=0.2$. The present values ($r_0$,$s_0$) is smallest compare
with previous cases of $\alpha$. For the parameteres ($n=1.2,
\alpha=0.2$), the evolutionary trajectory in open universe has a
different treatment in comparison with of flat and closed universe.
The today's values ($r_0,s_0$) for different contribution of spatial
curvature and model parameters $n$ and $\alpha$ are collected in table(1).\\

Finally, we do the $w-w^{\prime }$ analysis for interacting ADE
model in non flat universe. In this analysis, the standard $\Lambda
$CDM model corresponds to the fixed point
\{$w_{d}=-1$,$w_{d}^{\prime }=0$\} in the $w-w^{\prime }$ plane. The
evolution of $w_{d}$ and $w_{d}^{\prime }$ is given by
Eqs.(\ref{w_d}, \ref{wp_d}). In Fig.(4) the evolutionary
trajectories  in $w-w^{\prime }$ plane are shown for different model
parameters $n$ and $\alpha$ and also the contribution of different
spatial curvatures. The first arrow panels, show the evolutionary
trajectories for ADE model without interaction term ($\alpha=0.0$).
In this case, from Eqs.(\ref{w_d}, \ref{wp_d}) it is easy to see
that at the early time $w_d=-1.0$ and $w_d^{\prime}=0.0$. By
expanding the universe, the trajectories start from the point
($w=-1,w^{\prime}=0.0$), then $w_d$ increases and also
$w_d^{\prime}$ increases to a some maximum value, after that
decreases in $w-w^{\prime}$ plane. At the early time, the
contribution of spatial curvature can be neglected, and the
evolutionary trajectories start from the fixed point
($w=-1,w^{\prime}=0.0$) for any spatial curvature. The colore points
on the curves represent the present values of \{$w_d,w^{\prime}$\}.
 One can see that different spatial
curvatures of the universe result different evolutionary
trajectories in $w-w^{\prime}$ plane. The present $w_d$ is equal for
all spatial curvature models, but $w_d^{\prime}$ in closed universe
is largest compare with open and flat universe. The evolutionary
trajectory is also dependent on the model parameters $n$ and
$\alpha$. Different values of $n$ result the different trajectories.
The maximum of $w_d^{\prime}$ is smaller when $n$ is larger. Also,
the present value of $w_d$ is smaller and the present value of
$w_d^{\prime}$ is larger when the parameter $n$ is larger. In second
arrow panels, the trajectories are plotted for $\alpha=0.1$. In this
case, from Eqs.(\ref{w_d}, \ref{wp_d}), it is obvious that the
initial values of $w_d$ and $w_d^{\prime}$  at the early time are
$w_d=-1.1$ and $w_d^{\prime}=0.0$. In this case the present $w_d$
and $w_d^{\prime}$ is smaller compare with the previous case
($\alpha=0.0$). In third arrow panels, the evolutionary trajectories
are shown for $\alpha=0.2$. For the parameters ($n=0.9,
\alpha=0.2$), we see the different behavior of evolutionary
trajectories. The trajectories start form the point $w_d=0.1,
w_d^{\prime}=0.0$, then $w_d$ decreases, and $w_d^{\prime}$
decreases to a minimum value, after that increases. In the case of
$n=1.2, \alpha=0.2$, the trajectories in open universe is completely
different from flat and closed universe. Similar with the previous
cases, in the case of $\alpha=0.2$, the present values of
$w_d,w_d^{\prime}$ are larger when $n$ is larger. The present values
of $w_d,w_d^{\prime}$ for different model parameters $n$ and
$\alpha$ and spatial curvature $\Omega_k$, are presented in table
(2).

\section{Conclusion}
In this paper, the interacting ADE model has been extended in
a non flat universe. This extension can be summarized as:\\
(\emph{i}) we studied the evolutionary treatment of EoS parameter,
$w_d$, and the deceleration parameter, $q$, in different
contribution of spatial curvature. We showed that for any spatial
curvature, the ADE model without interaction term can not cross the
phantom divide. To achieve the phantom divide at any contribution of
curvature, the interaction term is needed. The ADE model gives the
decelerated expansion at the early time and then accelerated
expansion later. The transition from decelerated phase to
accelerated phase is dependent on parameters $n$ and $\alpha$ as
well as the type of spatial curvature of the universe. The universe
undergoes the accelerated expansion earlier, for larger values of
$n$ and $\alpha$. Also for the same values of $n$ and $\alpha$, the
accelerated phase in open universe tacks place sooner compare with
flat and closed universe.\\ (\emph{ii}) We performed the statefinder
diagnostic and $w-w^{\prime}$ analysis for the interacting ADE model
in non-flat universe. The statefinder diagnostic tool and
$w-w^{\prime}$ analysis are useful methods to discriminate the
various models of dark energy. Moreover, the present values of \{r,
s\} and \{$w$, $w^{\prime}$\}, if can be extracted from precise
observational data in a model-independent way, can be as a possible
discriminator for testing the cosmological models of dark energy. We
showed that the evolutionary trajectories in statefinder parameters
plane and $w-w^{\prime}$ plane are different for various types of
spatial curvatures. Also, the present values of \{r, s\} and \{$w$,
$w^{\prime}$\} are different for various spatial curvatures. Hence,
the statefinder diagram and $w-w^{\prime}$ analysis showed that the
contributions of the spatial curvature in the model can be diagnosed
out explicitly in this methods. We hope that the future
high-precision observations such as the SNAP-type experiment be
capable to determine the statefinder parameters precisely and
consequently single out the right cosmological models.\\

\newpage
\begin{table}
\begin{center}
\caption{The present values of the statefinder parameters ($r,s$)
for different model parameters $n$ and $\alpha$ in closed, flat and
open universe. \label{tab1}}
\begin{tabular}{crrrr}
\tableline
  $\Omega_{k0}=0.02$ (Closed universe)&$n=0.9$ & $n=1.2$ & $n=2.0$\\
\tableline
$\alpha=0.0$&(0.162,0.730)& (0.102,0.545)&(0.2770,0.3261) \\
$\alpha=0.1$&(0.029,0.660)&(0.003,0.506)&(0.217,0.308)\\
$\alpha=0.2$&(-0.102,0.614)&(-0.095,0.478)&( 0.158,0.294)\\
 \tableline
 $\Omega_{k0}=0.0$ (Flat universe)&$n=0.9$ & $n=1.2$ & $n=2.0$\\
\tableline
$\alpha=0.0$&(0.133,0.718)&(0.070,0.538)&(0.240,0.323)\\
$\alpha=0.1$&(0.002,0.650)&(-0.028, 0.500)&(0.181,0.305)\\
$\alpha=0.2$&(-0.130, 0.606)&(-0.127,0.473)&(0.121,0.292) \\
\tableline
  $\Omega_{k0}=-0.02$ (Open universe)&$n=0.9$ & $n=1.2$ & $n=2.0$\\
\tableline
$\alpha=0.0$&(0.109,0.707)&(0.042,0.533)&(0.2091,0.3208) \\
$\alpha=0.1$&(-0.022,0.642)&(-0.056, 0.496)&(0.149,0.303)\\
$\alpha=0.2$&(-0.154,0.599)&(-0.155,0.469)&(0.090,0.290) \\
\tableline
\end{tabular}
\end{center}
\end{table}

\begin{table}
\begin{center}
\caption{The present values of the parameters ($w,w^{\prime}$) for
different model parameters $n$ and $\alpha$ in closed, flat and open
universe. \label{tab2}}
\begin{tabular}{crrrr}
\tableline
  $\Omega_{k0}=0.02$ (Closed universe)&$n=0.9$ & $n=1.2$ & $n=2.0$\\
\tableline
$\alpha=0.0$&( -0.367,0.107)& (-0.525,0.111)&(-0.7152,0.088) \\
$\alpha=0.1$&( -0.467, 0.038)&(-0.625,0.059)&(-0.815,0.057)\\
$\alpha=0.2$&(-0.567,-0.030)&(-0.725, 0.007)&(-0.915,0.026)\\
 \tableline
 $\Omega_{k0}=0.0$ (Flat universe)&$n=0.9$ & $n=1.2$ & $n=2.0$\\
\tableline $\alpha=0.0$&(-0.367,0.094)&(-0.525,0.101)&(-0.715,0.082)\\
$\alpha=0.1$&(-0.467,0.024)&(-0.625,0.049)&(-0.815,0.051)\\
$\alpha=0.2$&(-0.567,-0.044)&(-0.725,-0.003)&(-0.915,0.020) \\
\tableline
  $\Omega_{k0}=-0.02$ (Open universe)&$n=0.9$ & $n=1.2$ & $n=2.0$\\
\tableline
$\alpha=0.0$&(-0.367, 0.082)&(-0.525,0.092)&(-0.715,0.077) \\
$\alpha=0.1$&(-0.467,0.013)&(-0.625,0.040)&(-0.815,0.046)\\
$\alpha=0.2$&(-0.567,-0.056)&(-0.725,-0.011)&(-0.915,0.014) \\
\tableline
\end{tabular}
\end{center}
\end{table}

\newpage
\begin{center}
\begin{figure}[!htb]
\includegraphics[width=5cm]{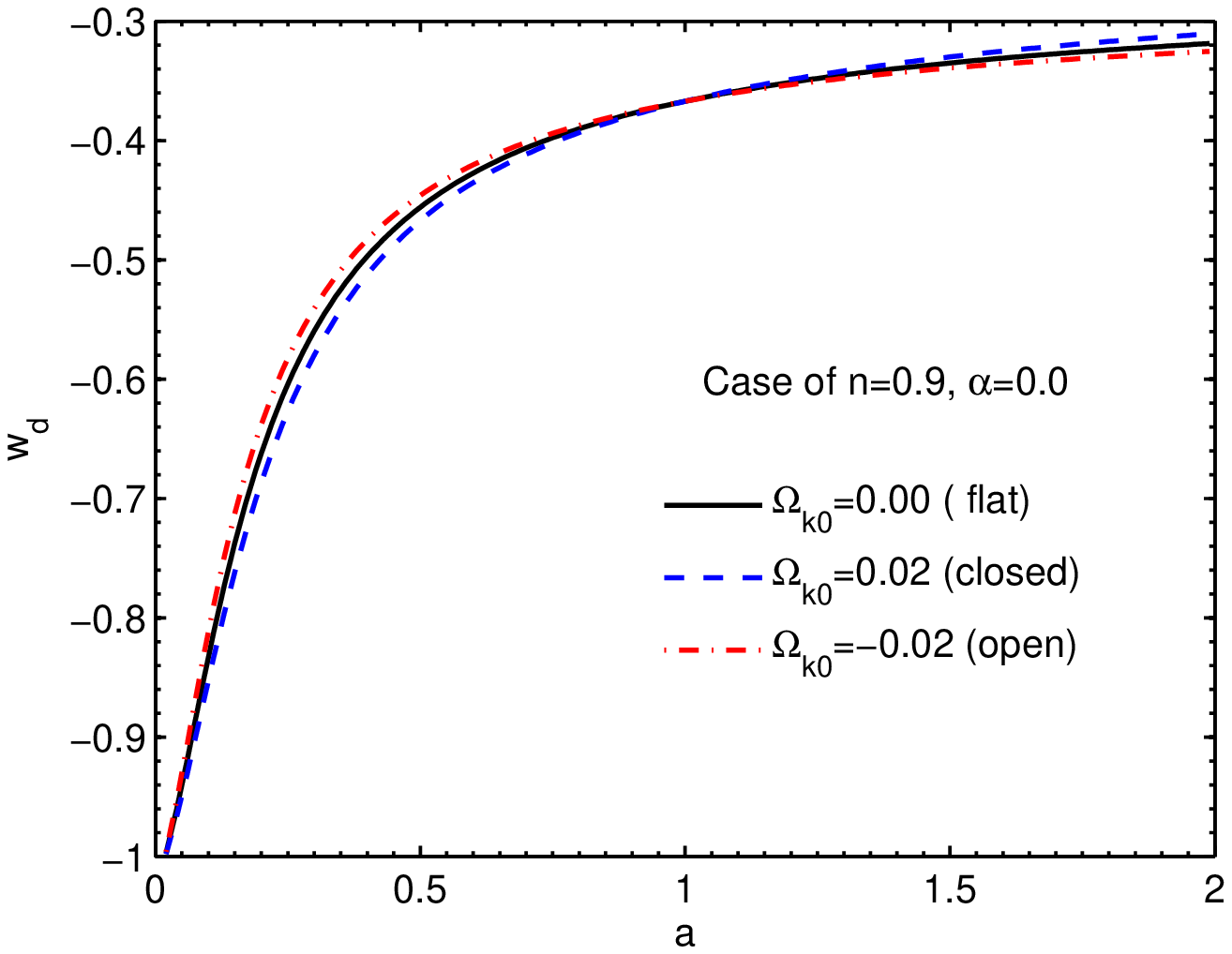} \includegraphics[width=5cm]{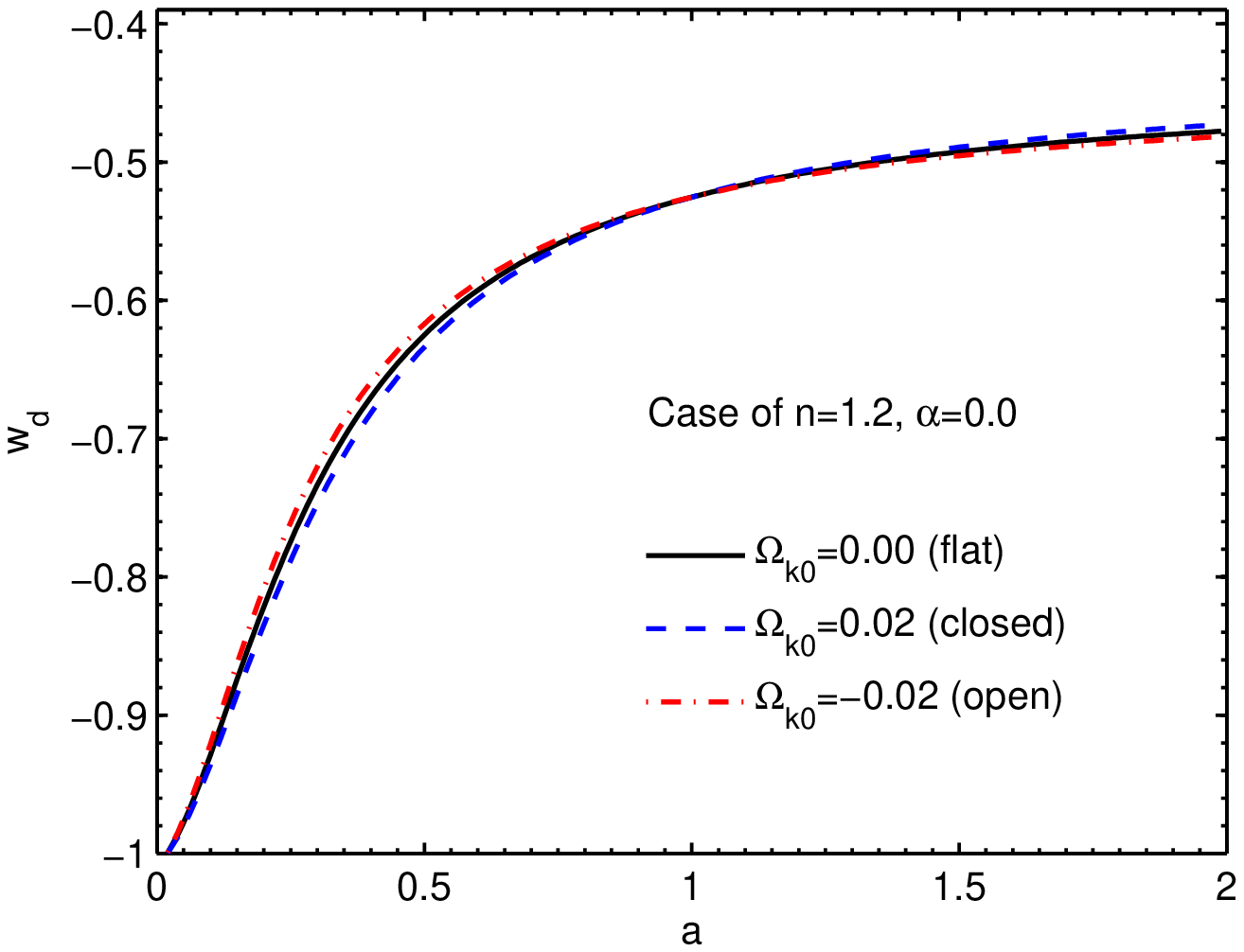} %
\includegraphics[width=5cm]{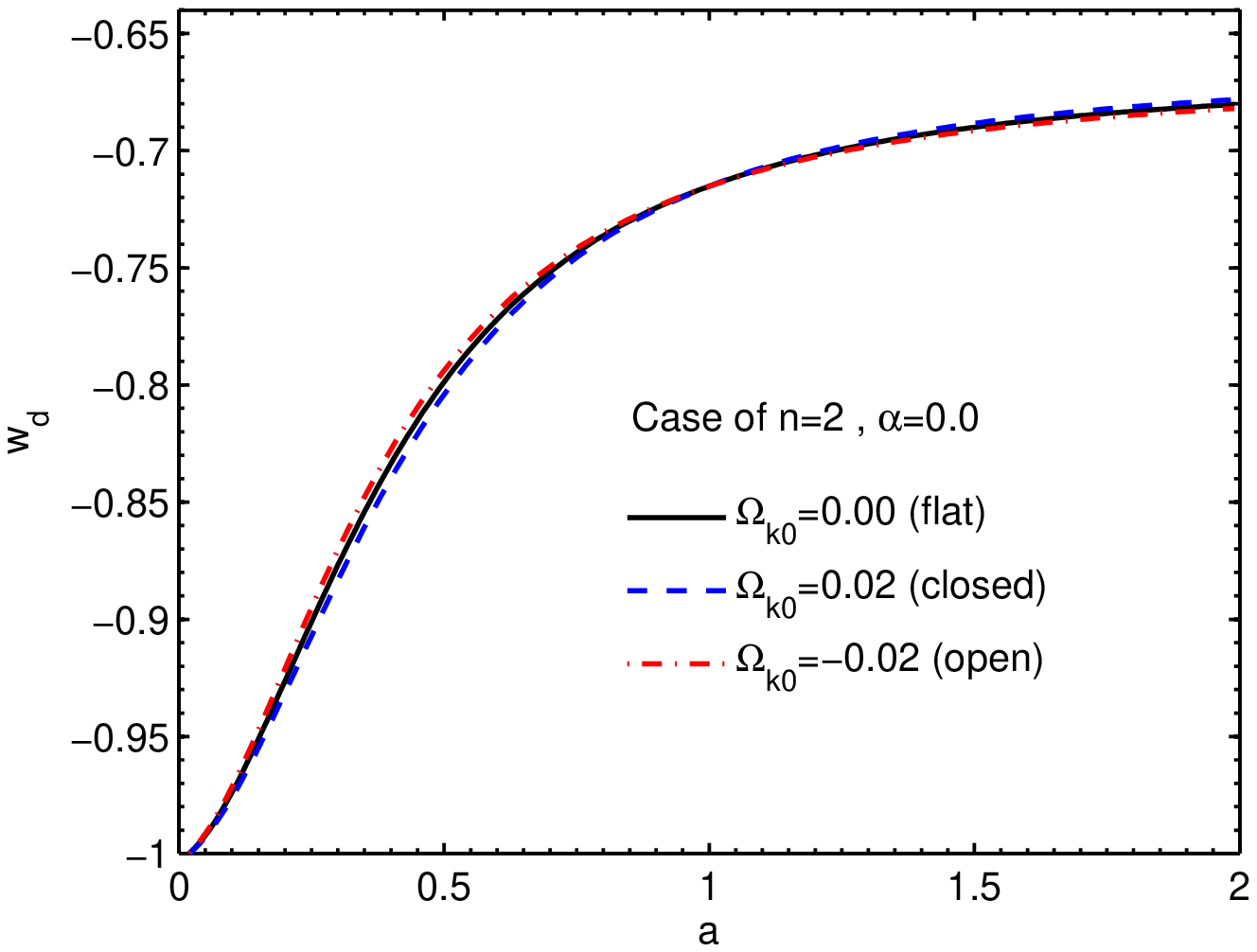} \includegraphics[width=5cm]{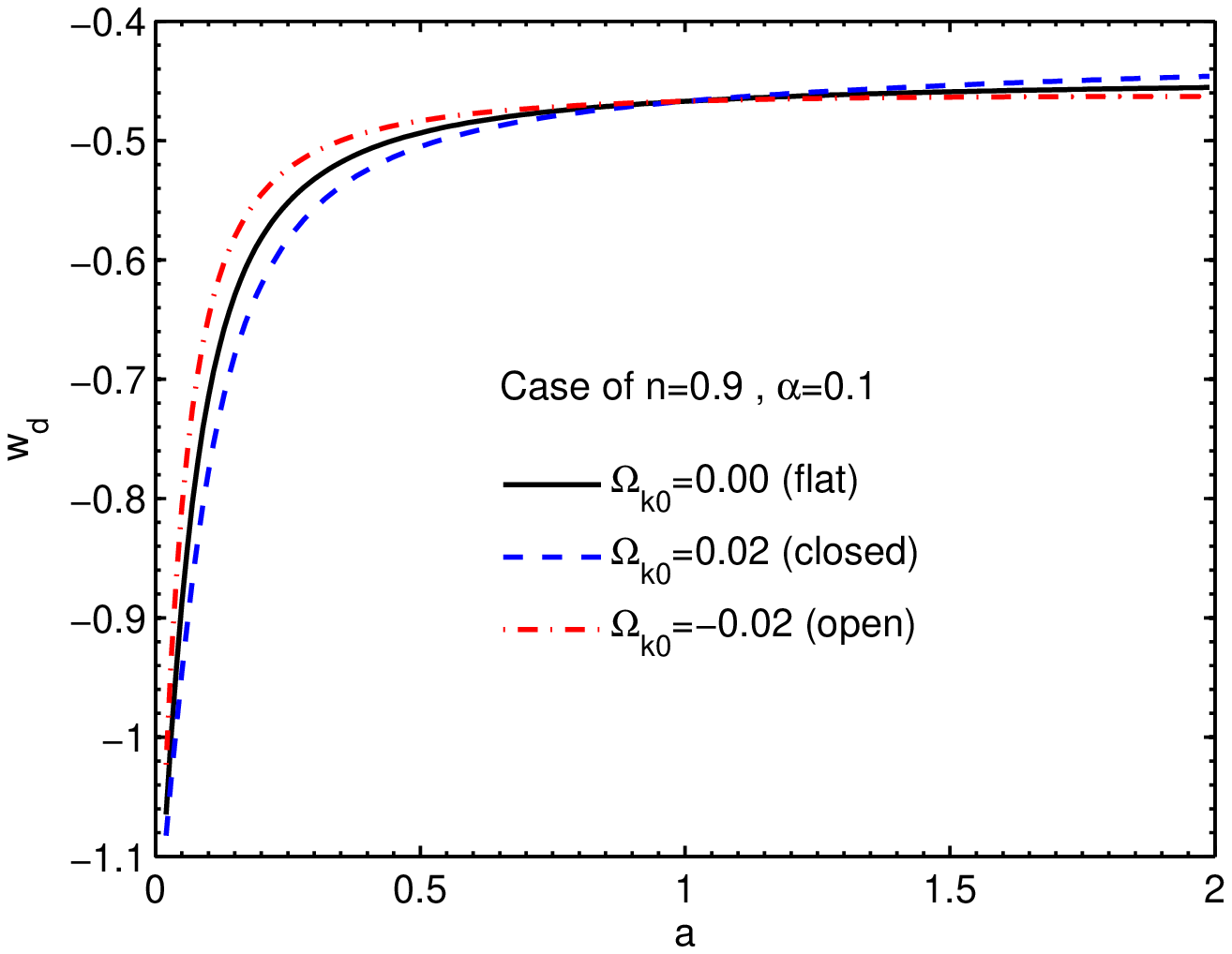} %
\includegraphics[width=5cm]{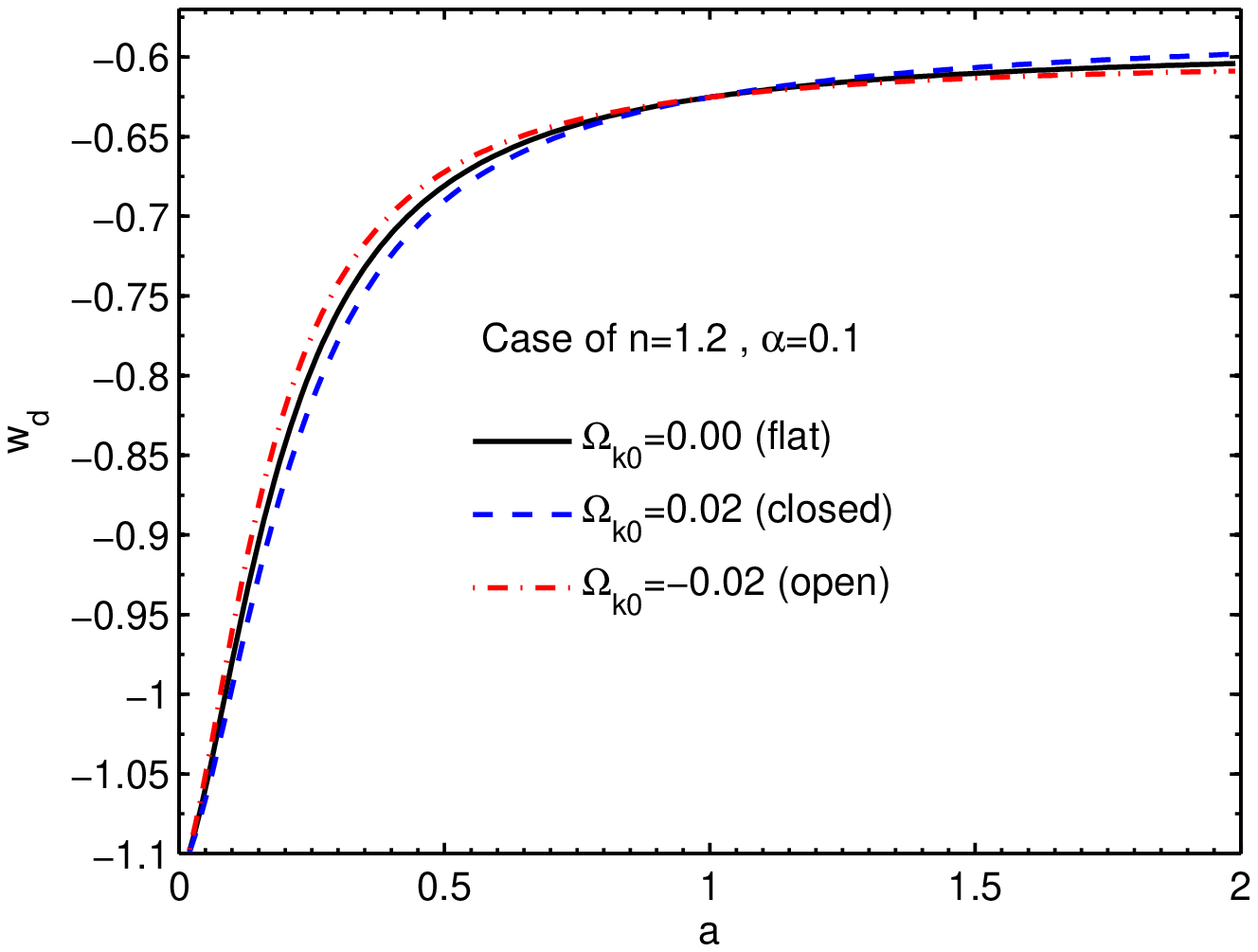} \includegraphics[width=5cm]{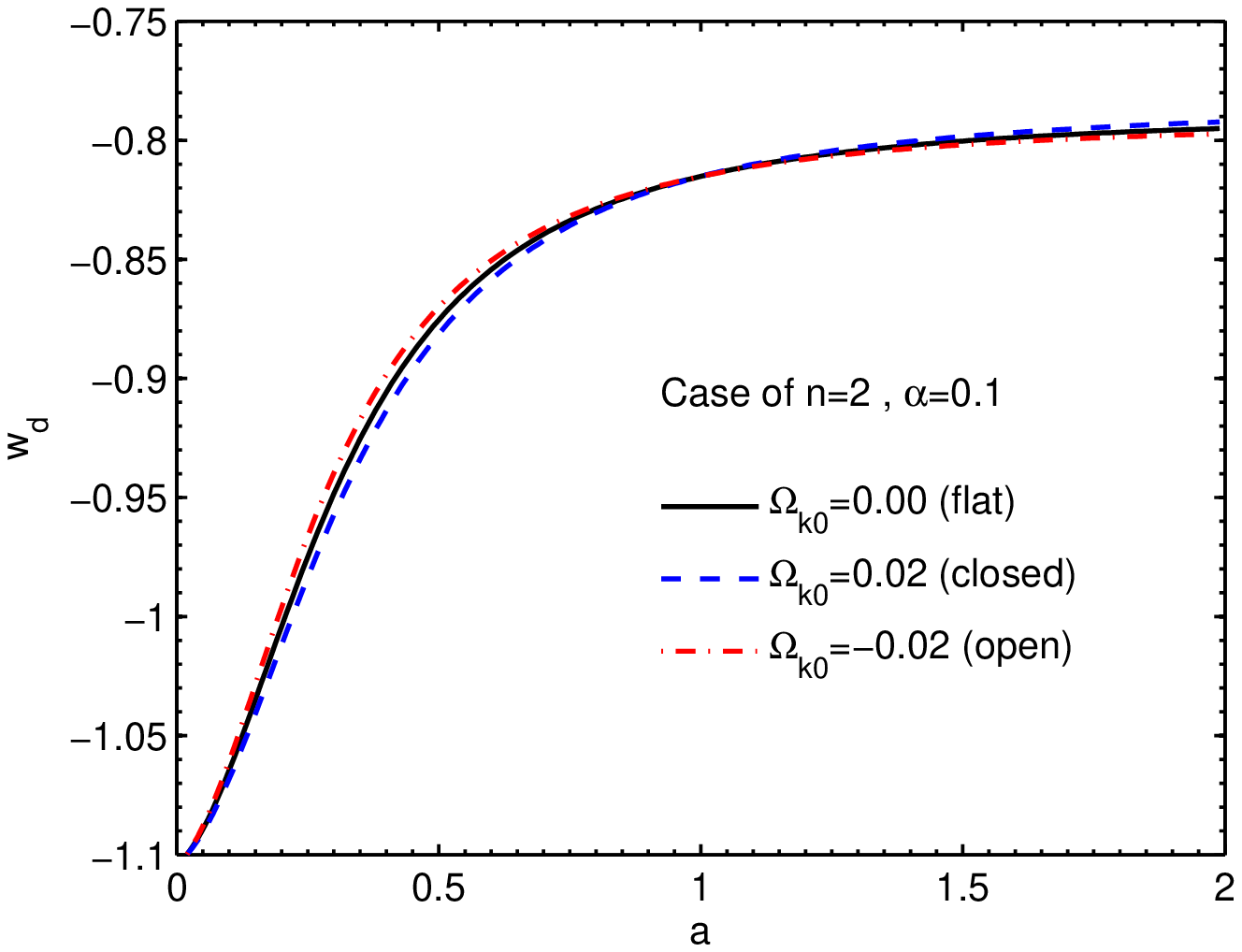} %
\includegraphics[width=5cm]{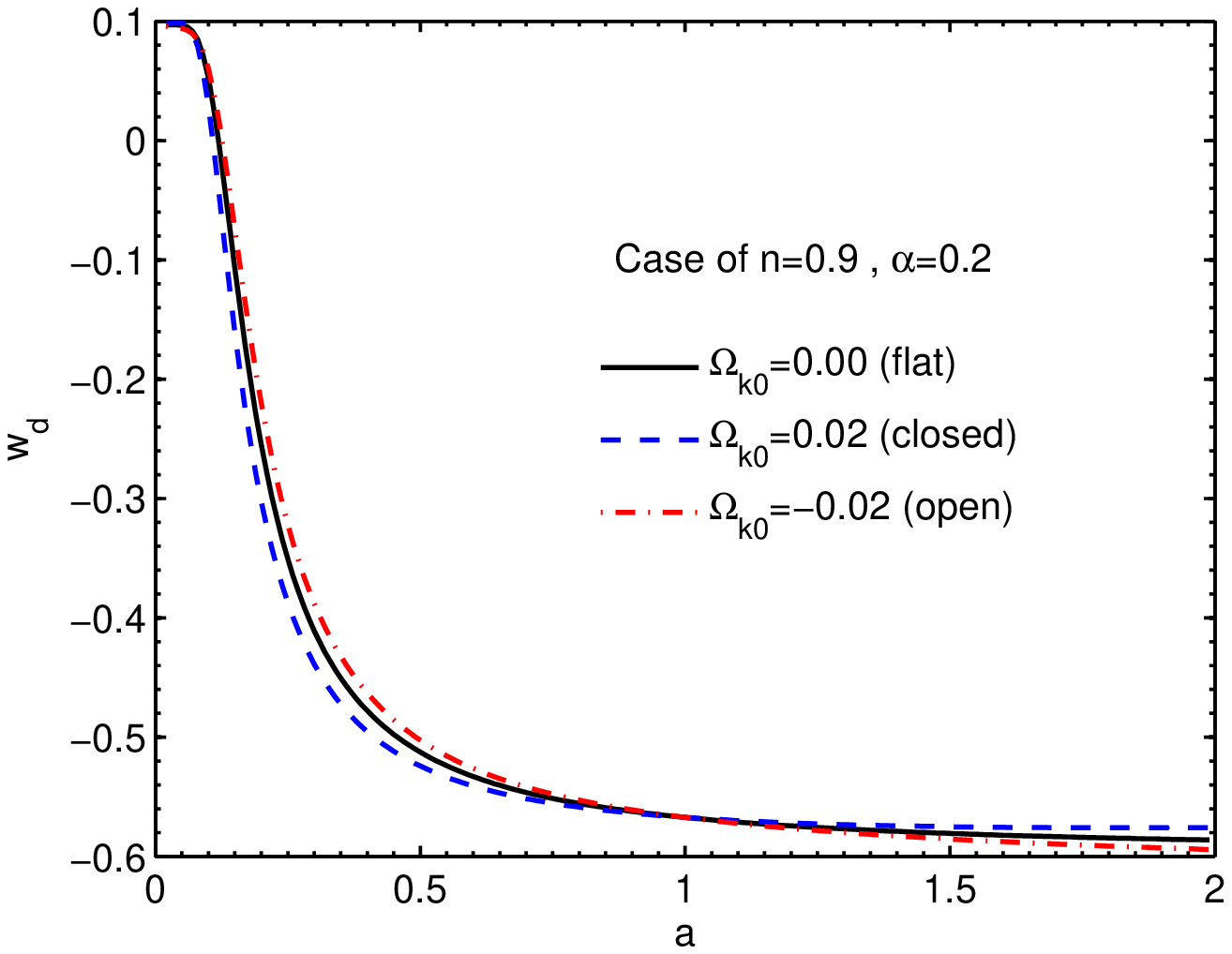} \includegraphics[width=5cm]{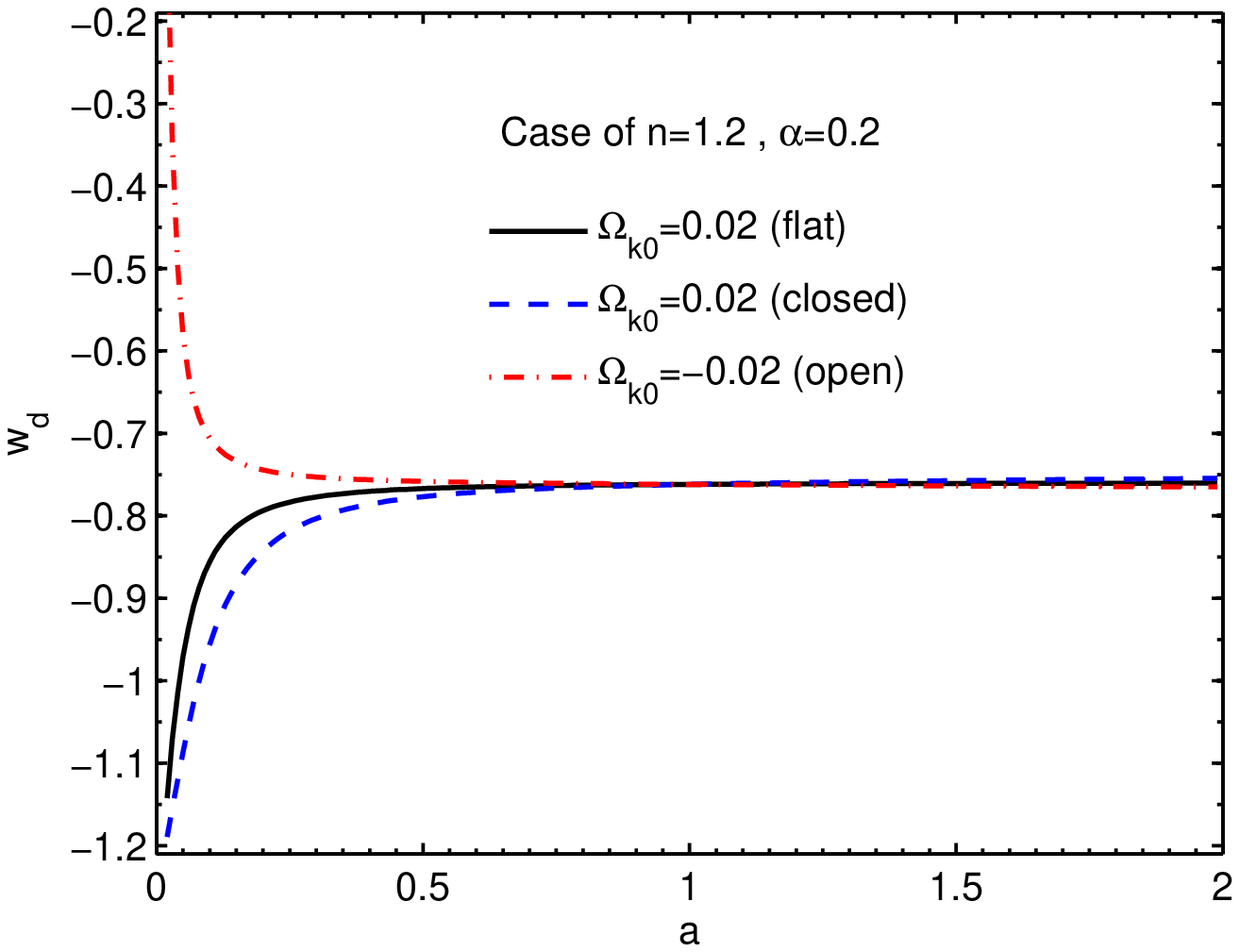}%
\includegraphics[width=5cm]{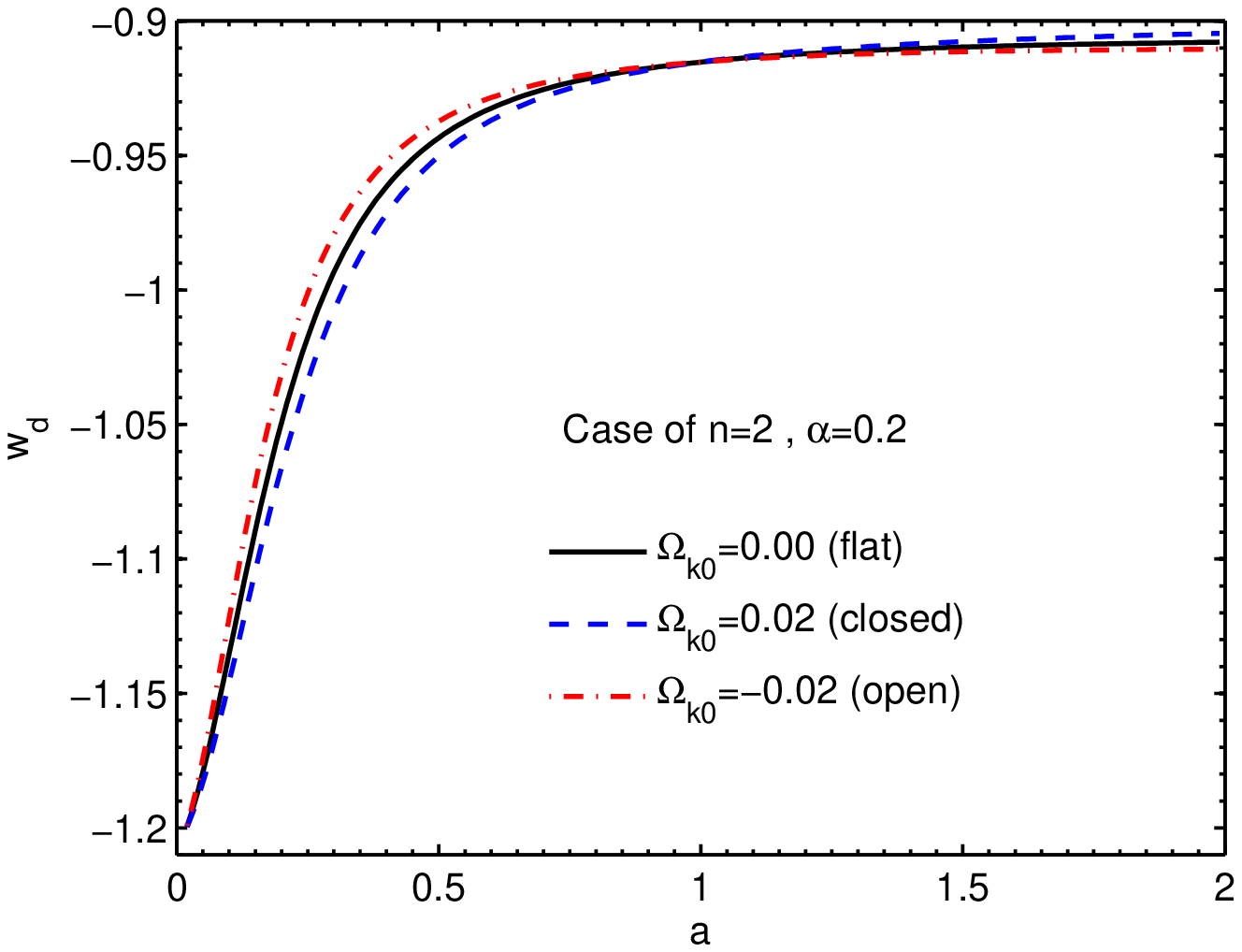}
\caption{The evolution of EoS parameter, $w_d$, versus of $a$ for
different model parameters $n$ and $\alpha$ in different closed,
flat and open universe. In first arrow, in the absence of
interaction between dark matter and dark energy, the phantom divide
can not be achieved. In second and third arrows, in the presence of
interaction, the phantom divide can be achieved.\\[0pt]}
\end{figure}
\end{center}
\newpage
\begin{center}
\begin{figure}[!htb]
\includegraphics[width=5cm]{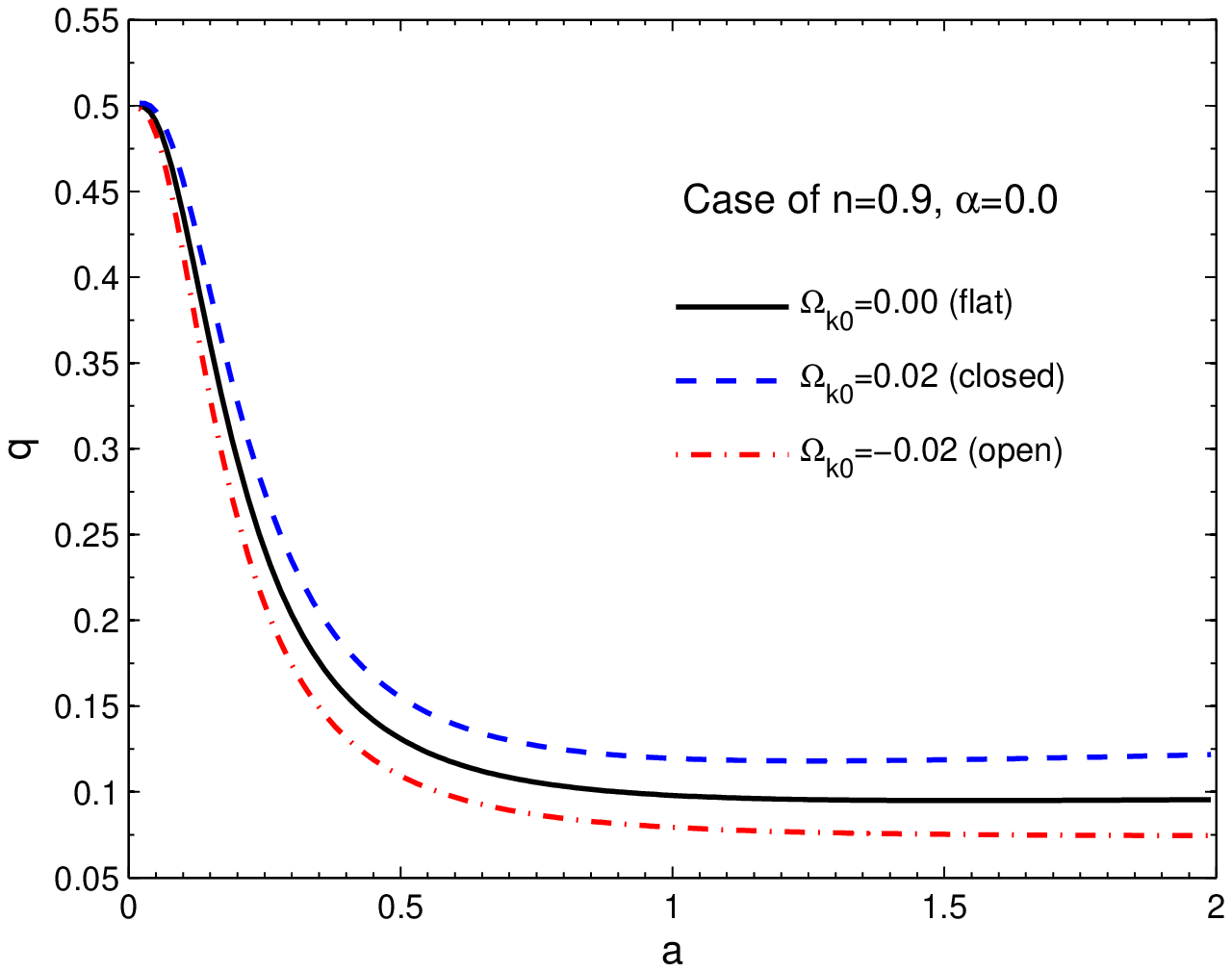} \includegraphics[width=5cm]{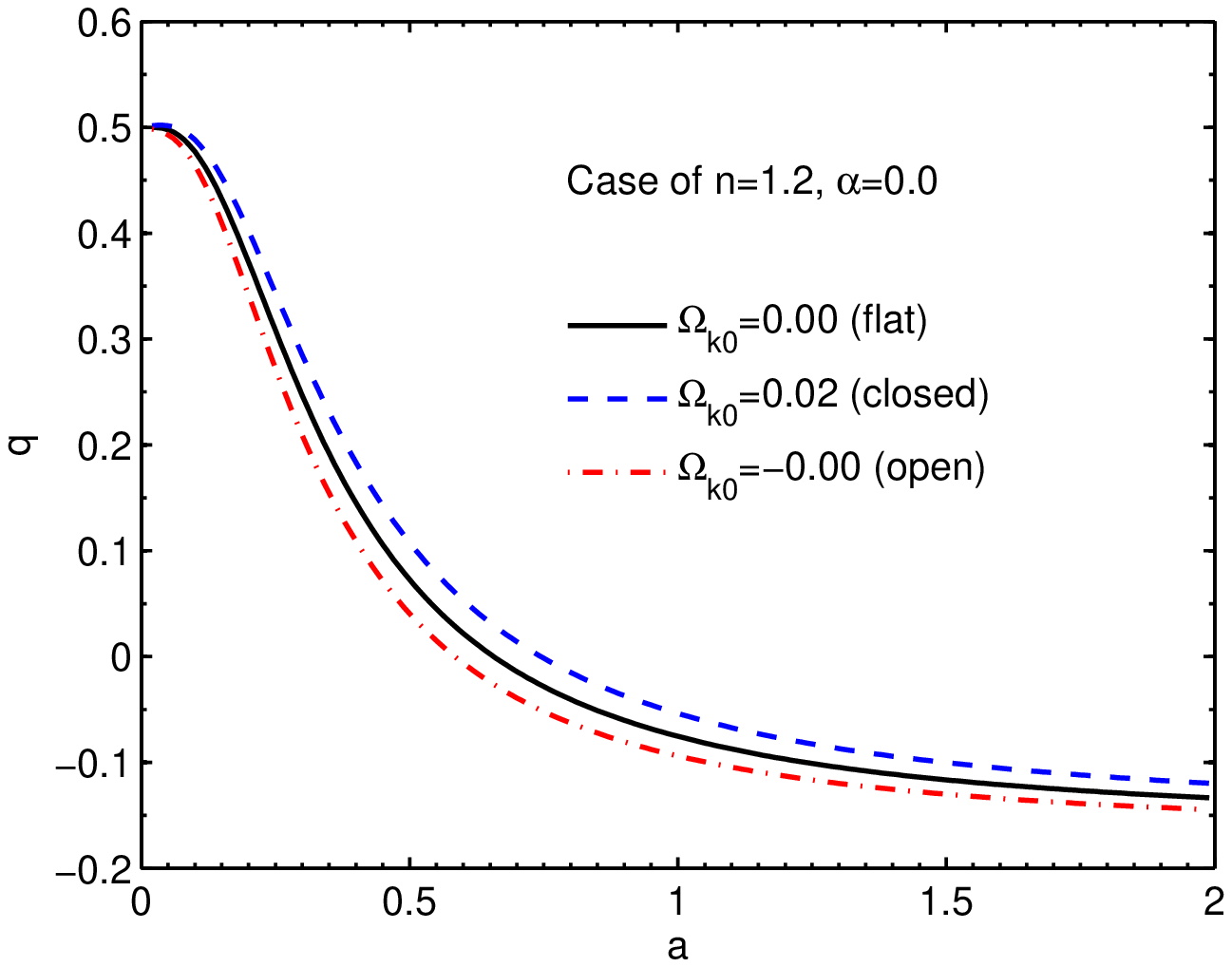} %
\includegraphics[width=5cm]{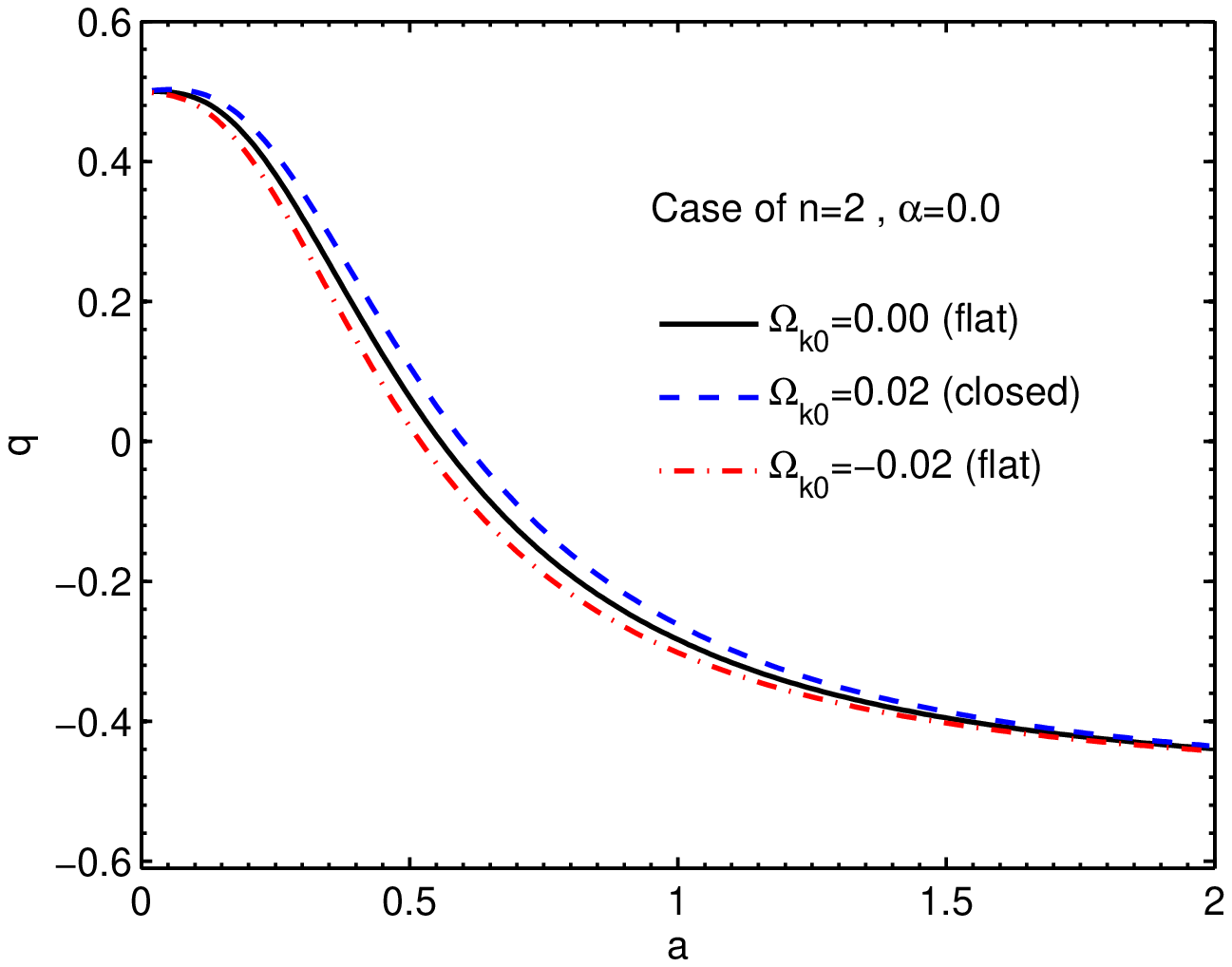} \includegraphics[width=5cm]{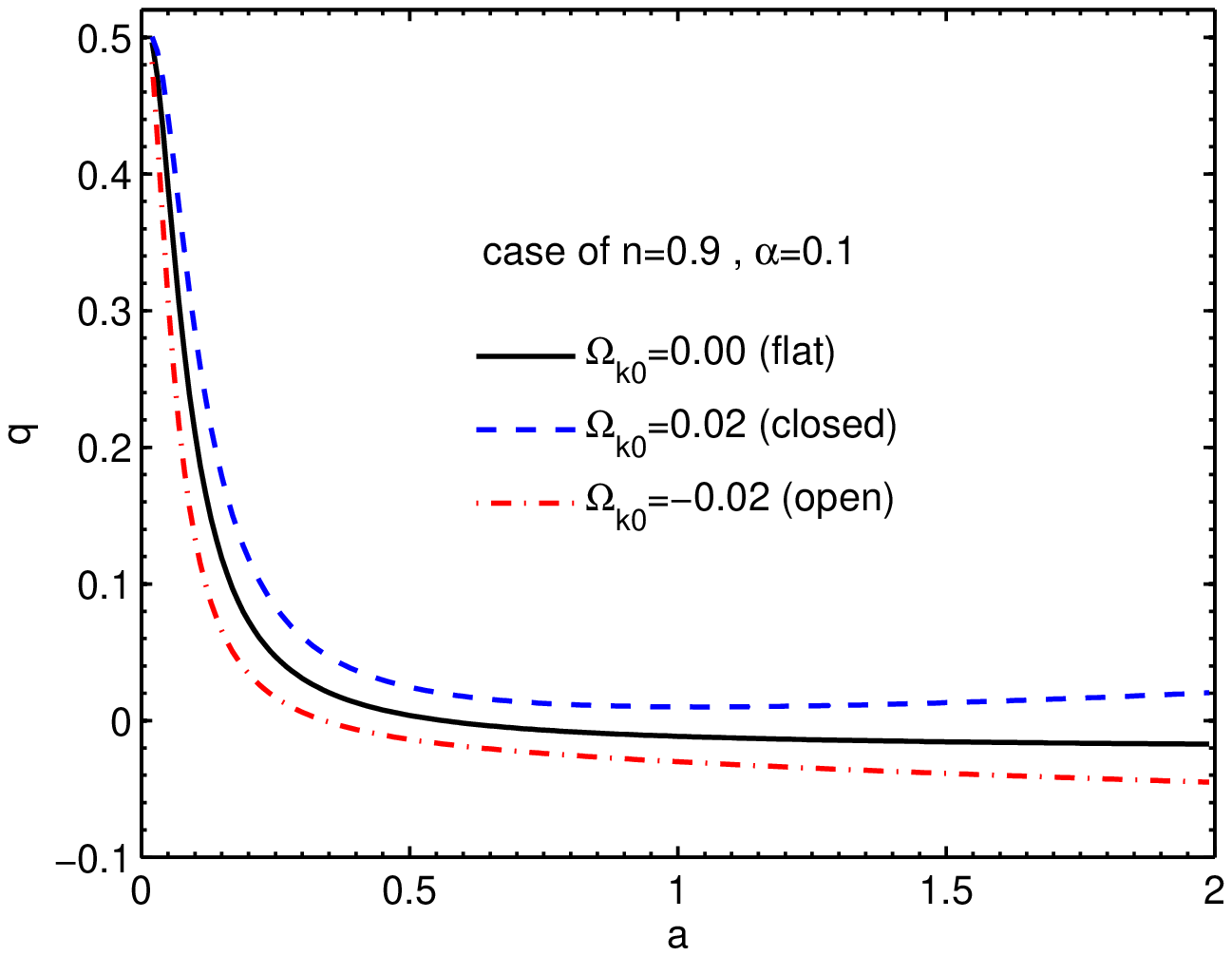} %
\includegraphics[width=5cm]{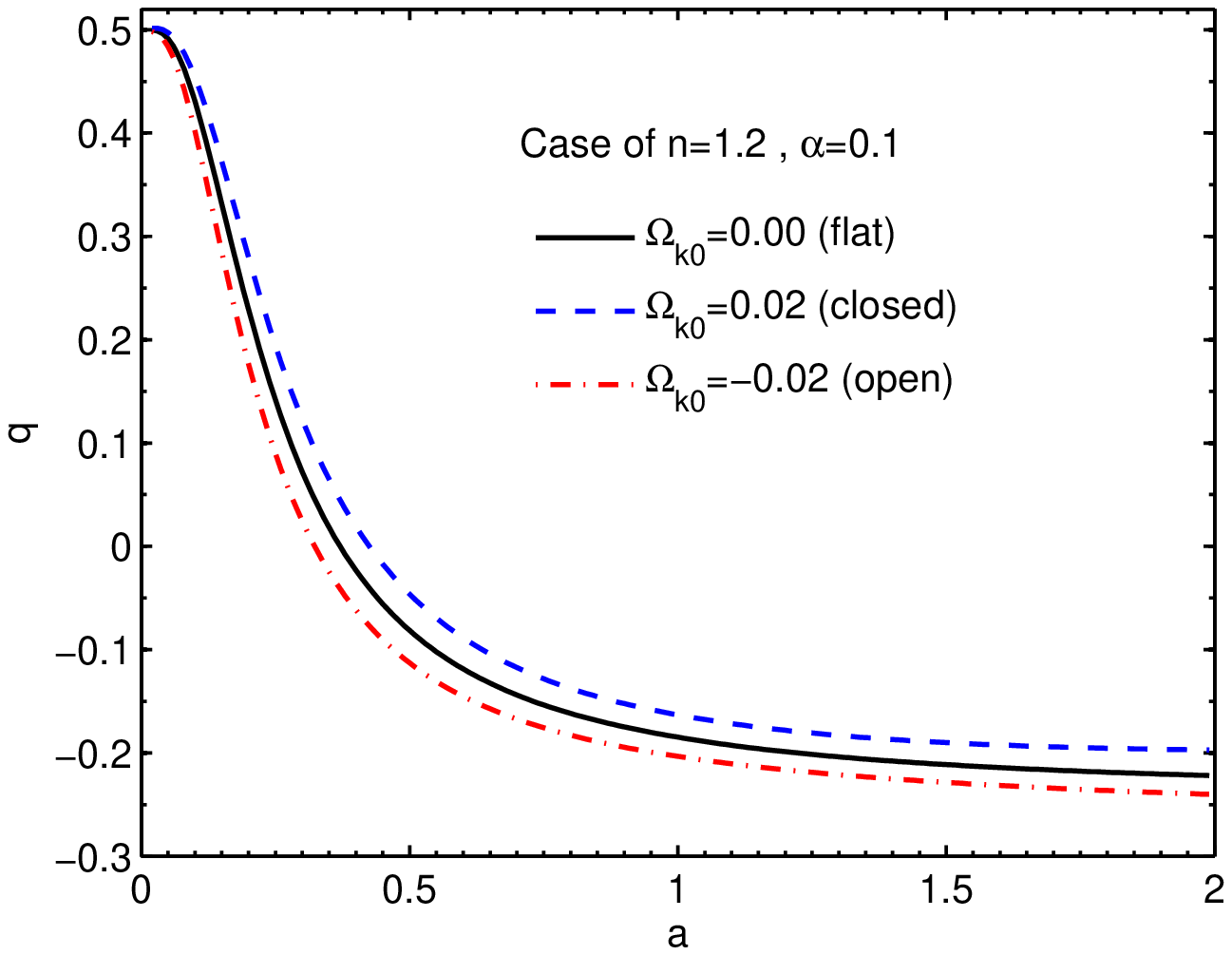} \includegraphics[width=5cm]{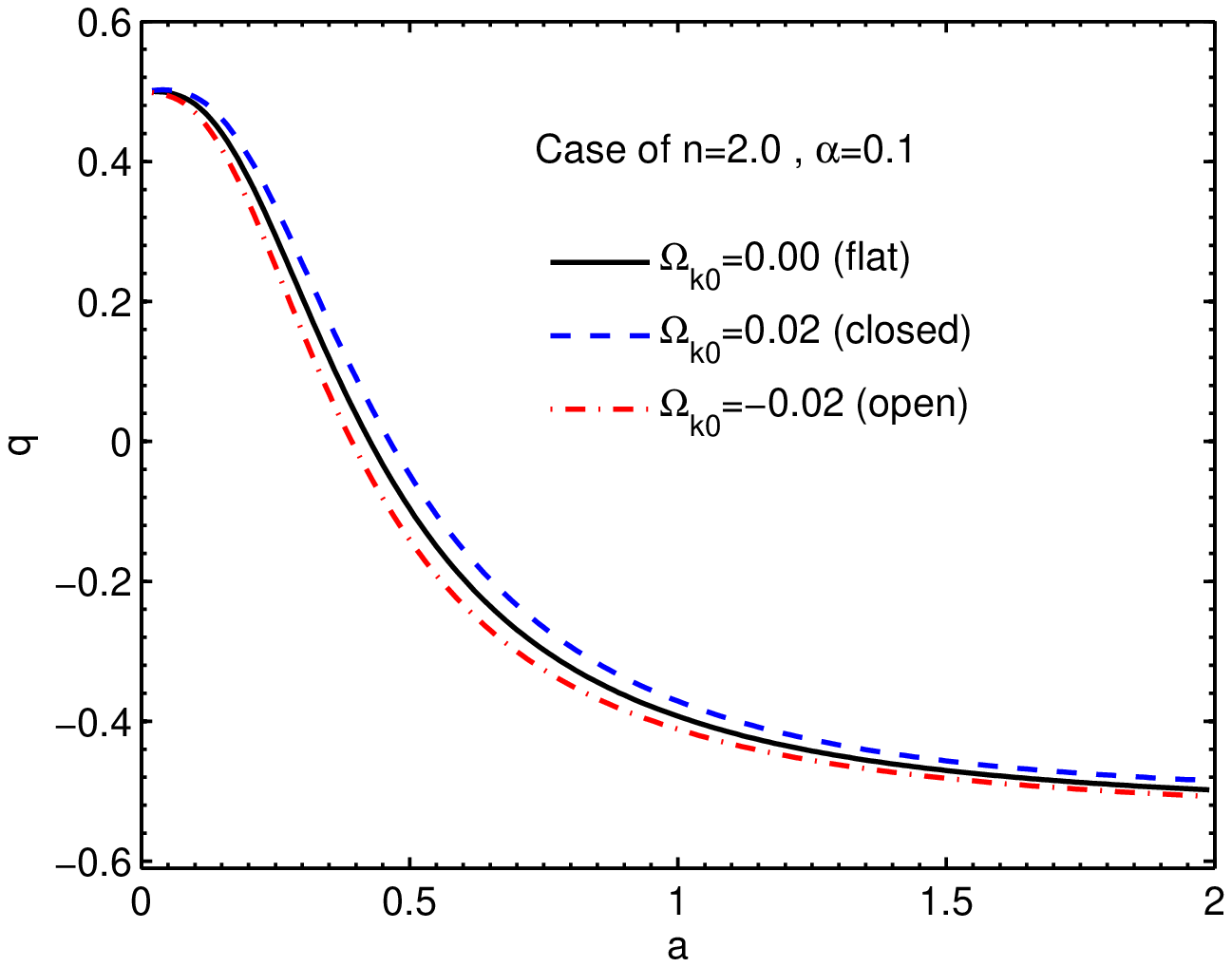} %
\includegraphics[width=5cm]{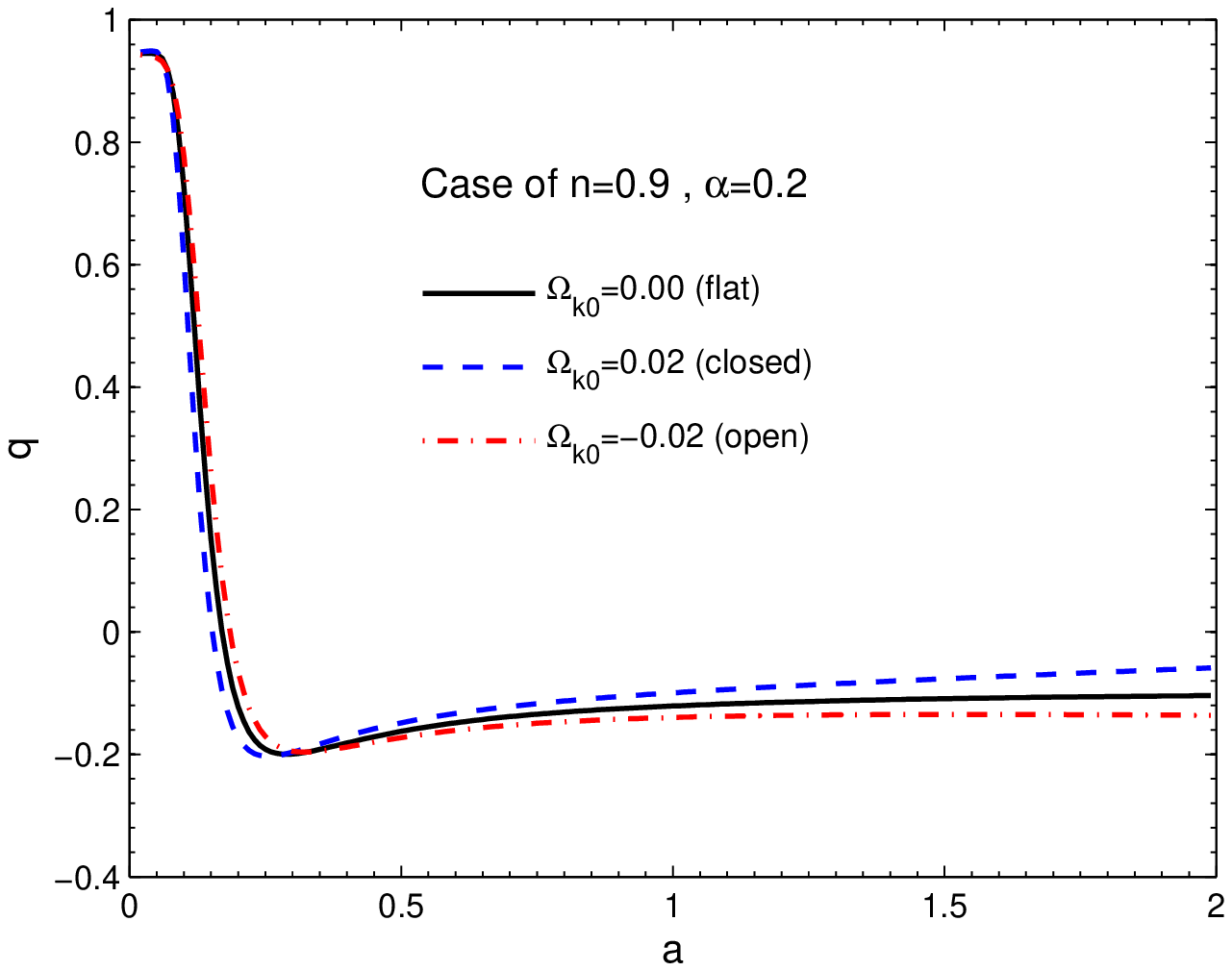} \includegraphics[width=5cm]{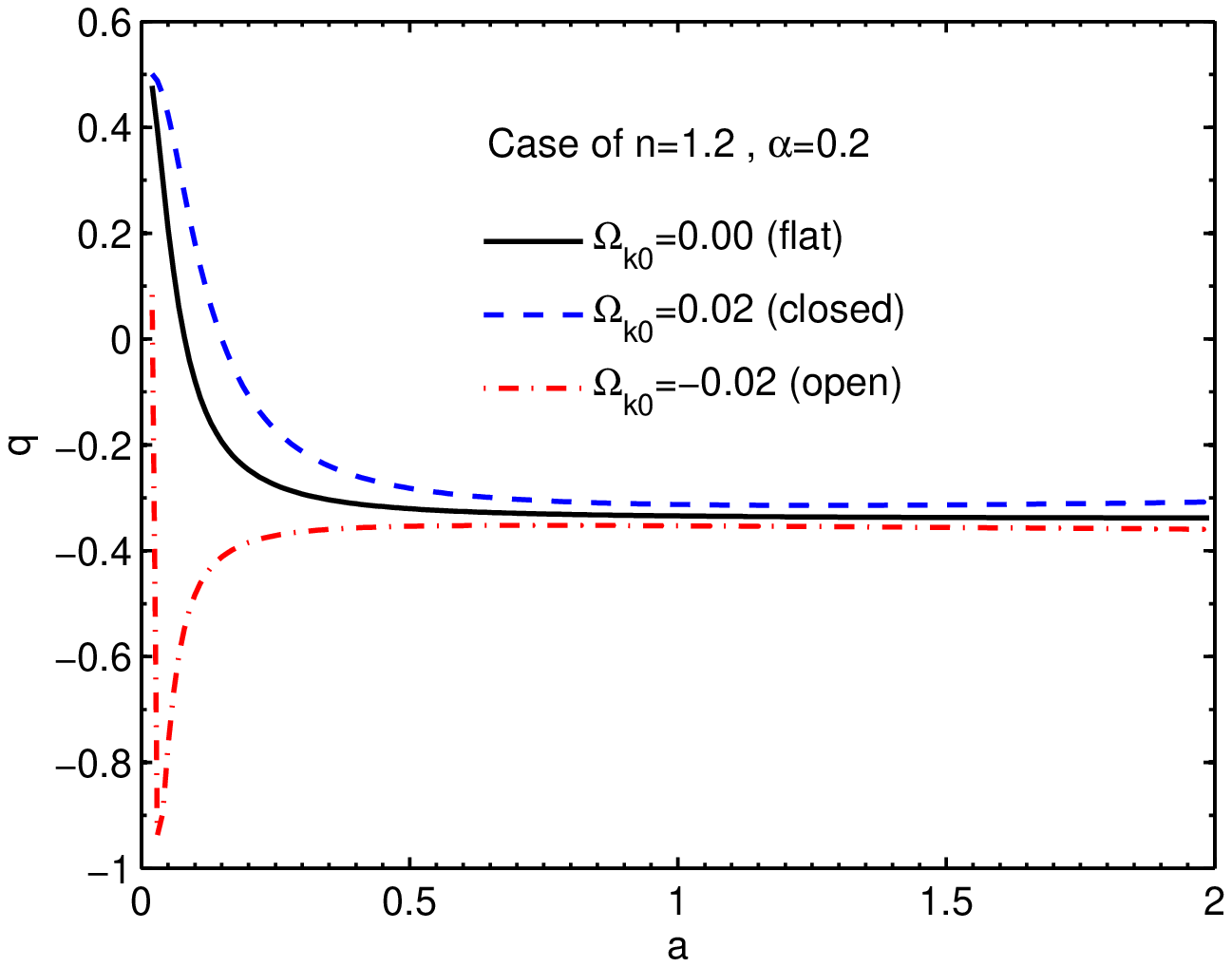}%
\includegraphics[width=5cm]{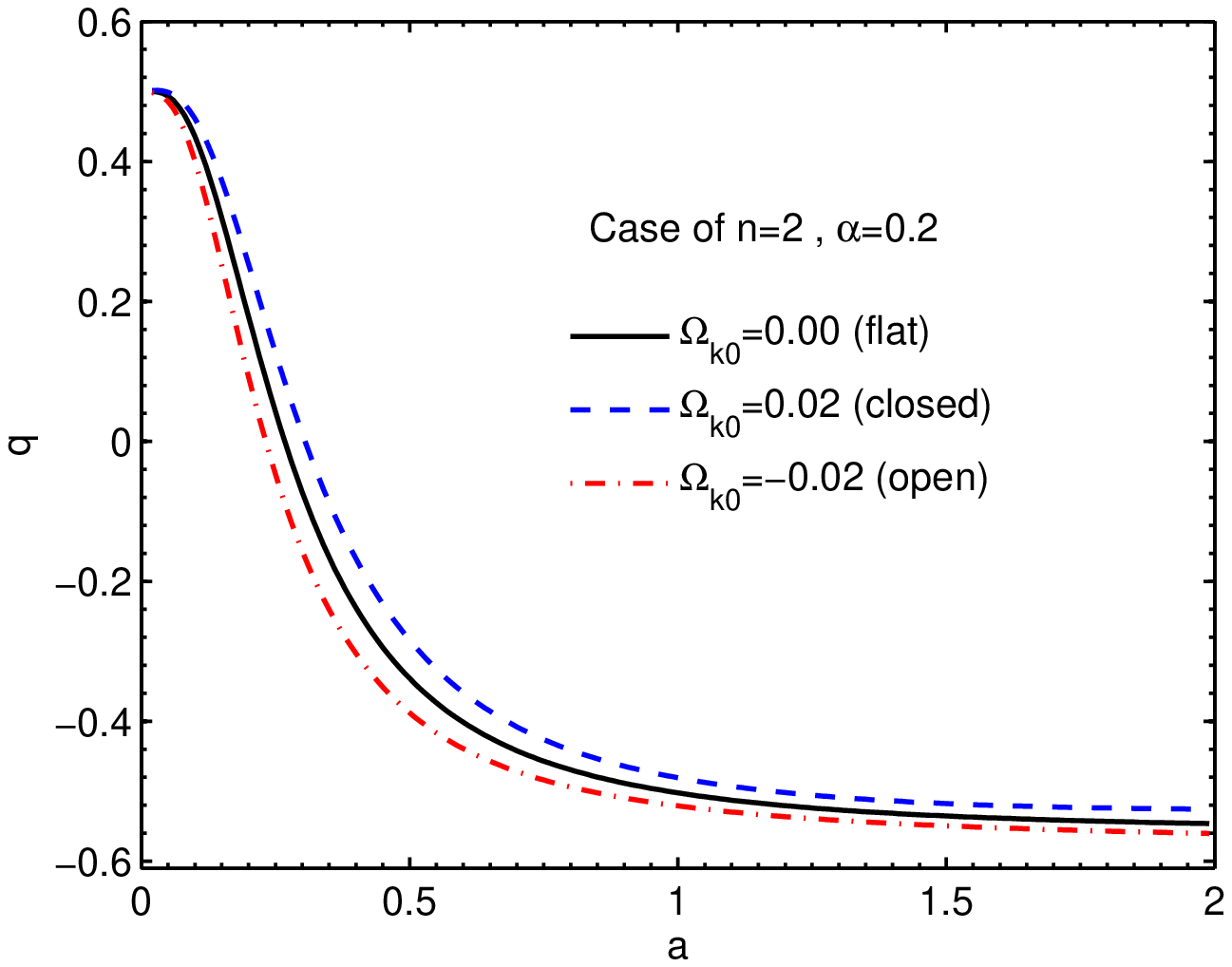}%
\caption{The evolution of deceleration parameter, $q$, versus of $a$
for different model parameters $n$ and $\alpha$ in different closed,
open flat and open universe. In first arrow the ADE model is assumed without
interaction term between dark components of the universe, while in the second and third
arrows the interaction term is included. The universe undergoes accelerated expansion earlier,
for larger values of $n$ and $\alpha$. Also, for the same values of parameters $n$ and $\alpha$
the transition from decelerated to the accelerated expansion occurs sooner in open universe. \\[0pt]}
\end{figure}
\end{center}
\newpage
\begin{center}
\begin{figure}[!htb]
\includegraphics[width=5cm]{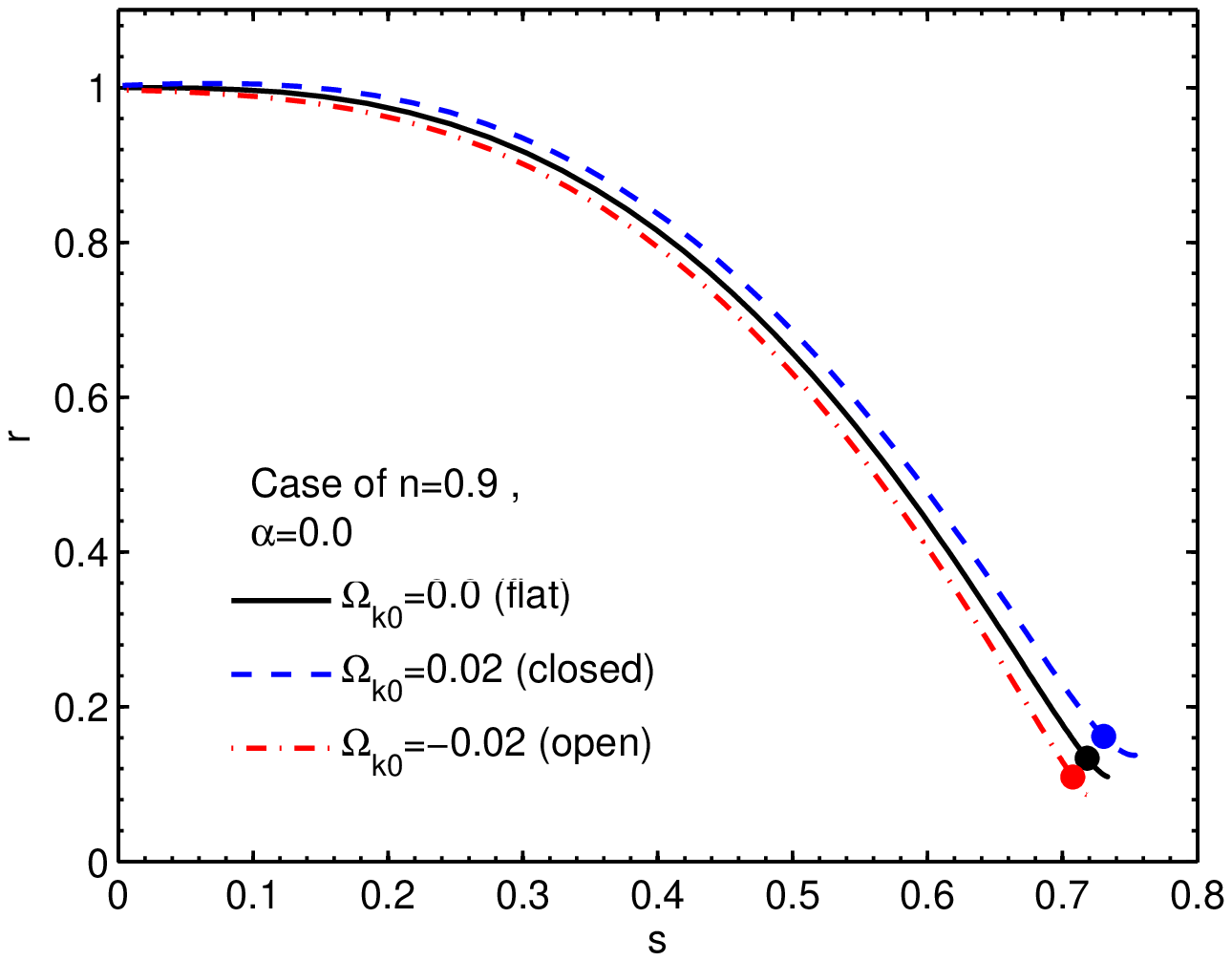} \includegraphics[width=5cm]{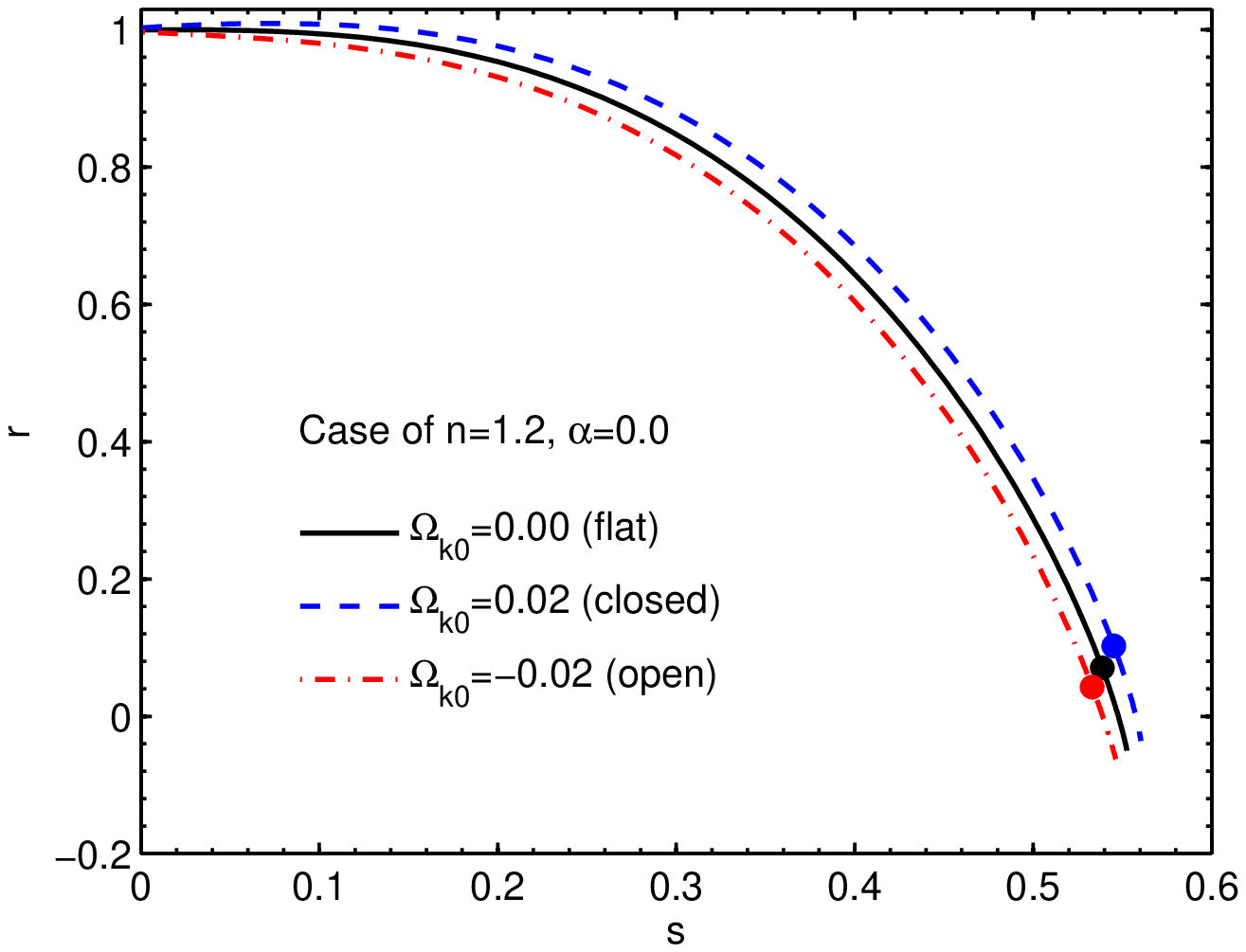} %
\includegraphics[width=5cm]{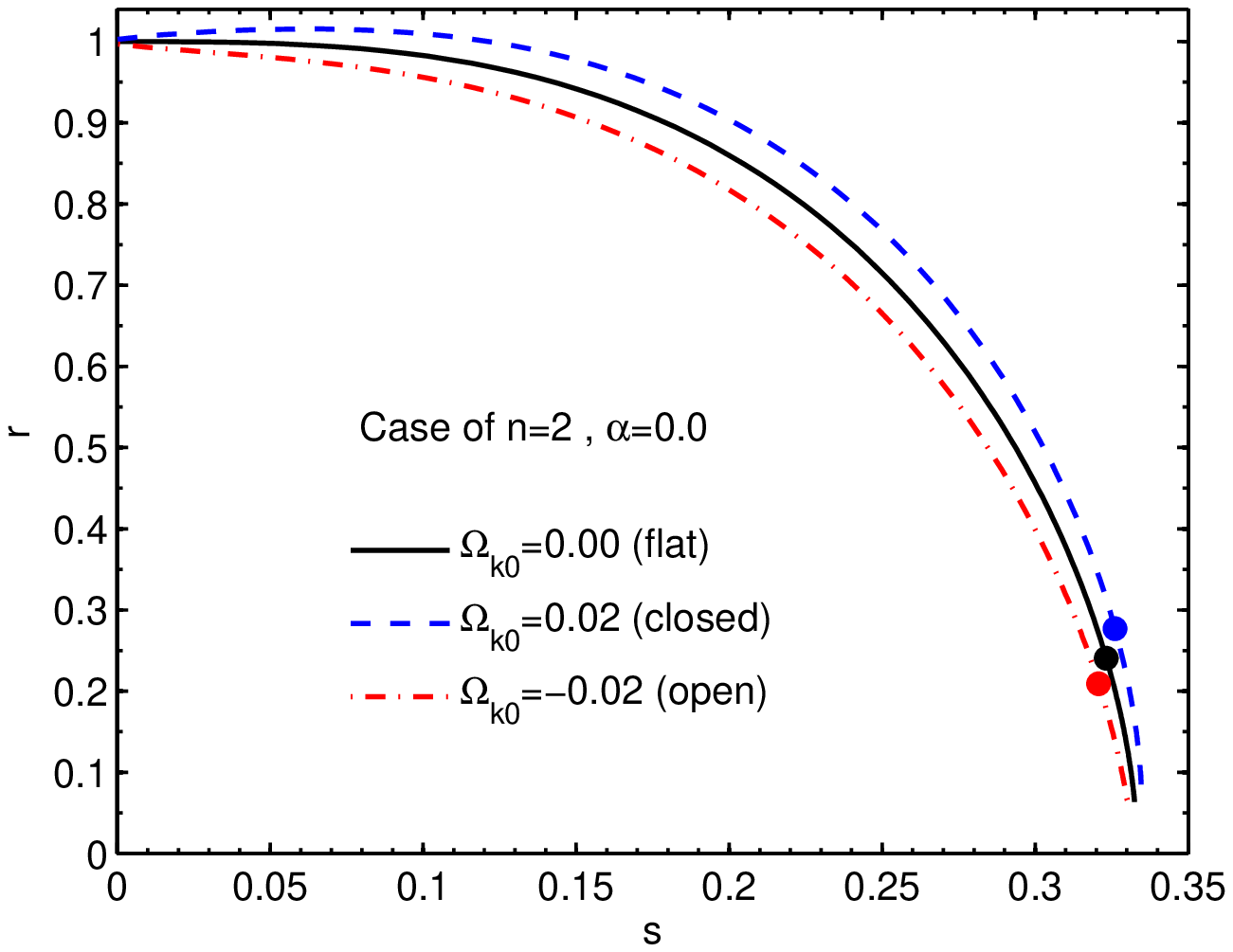} \includegraphics[width=5cm]{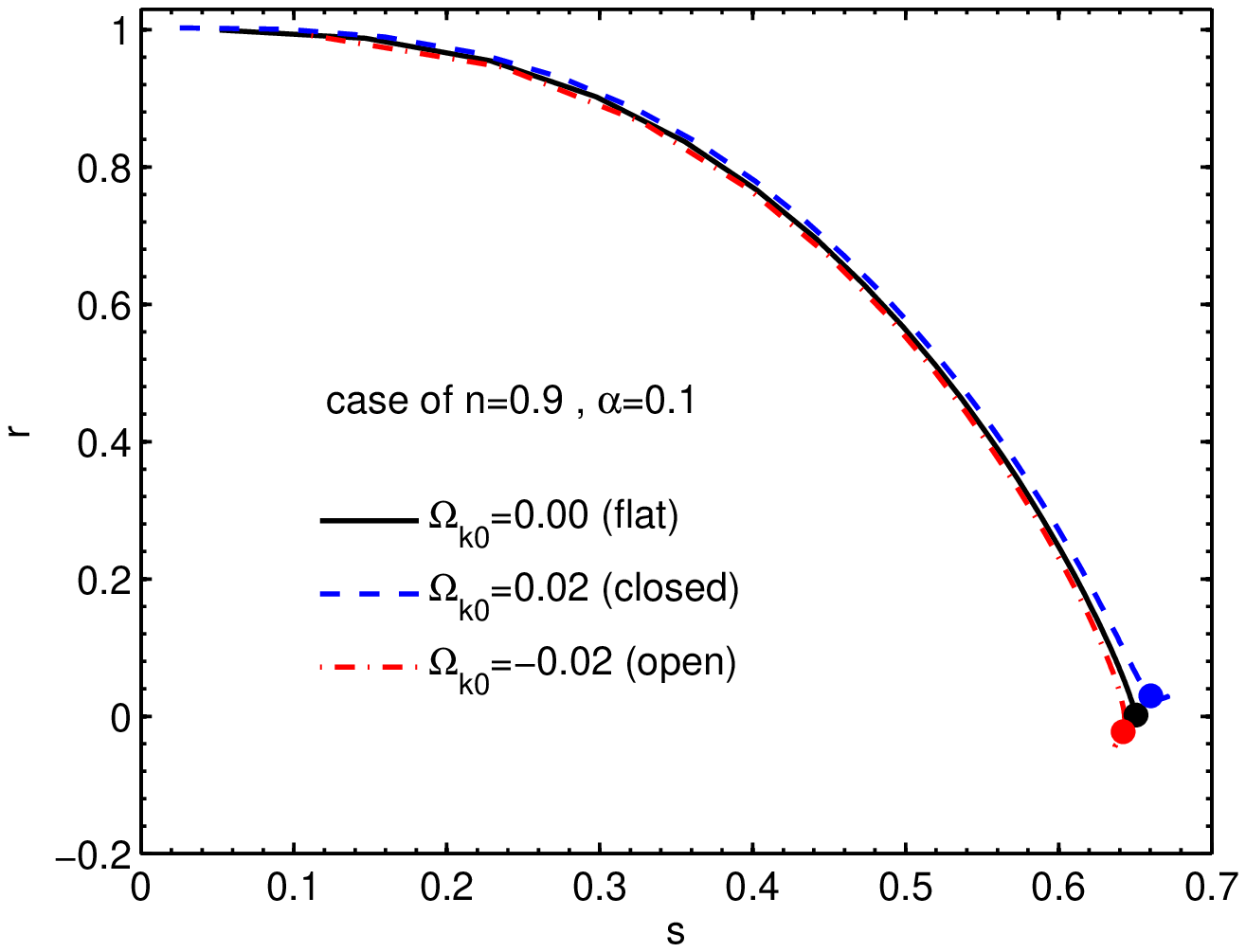} %
\includegraphics[width=5cm]{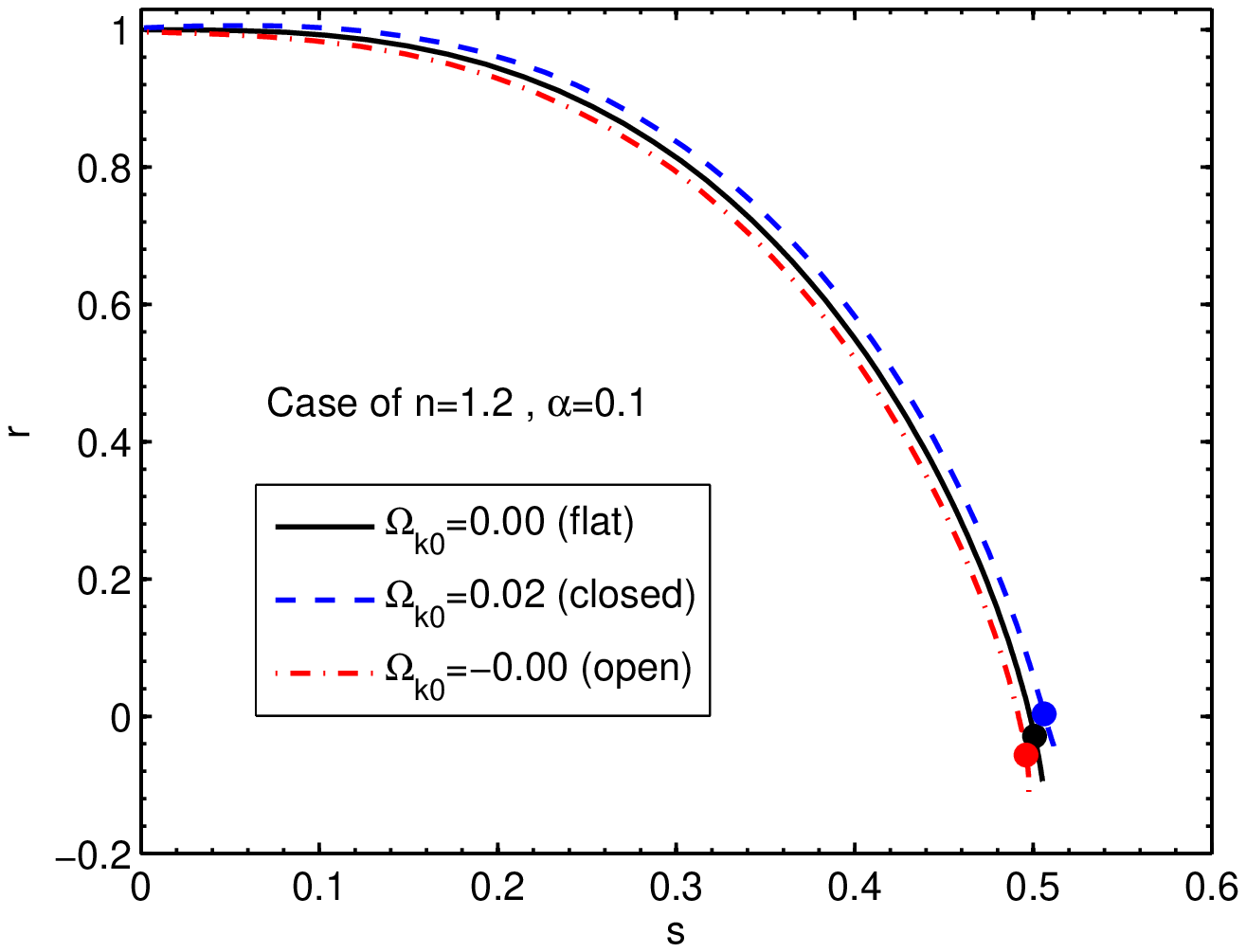} \includegraphics[width=5cm]{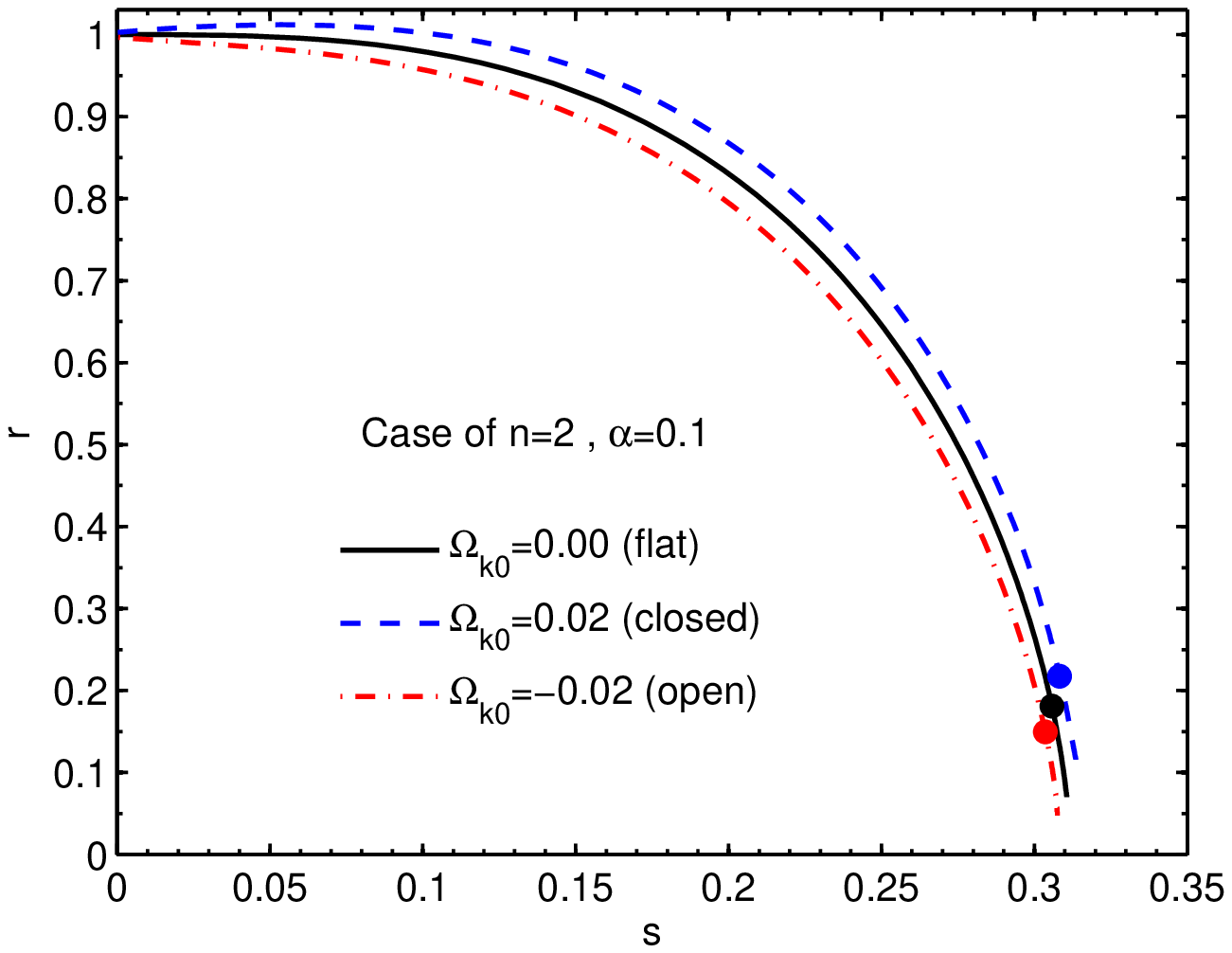} %
\includegraphics[width=5cm]{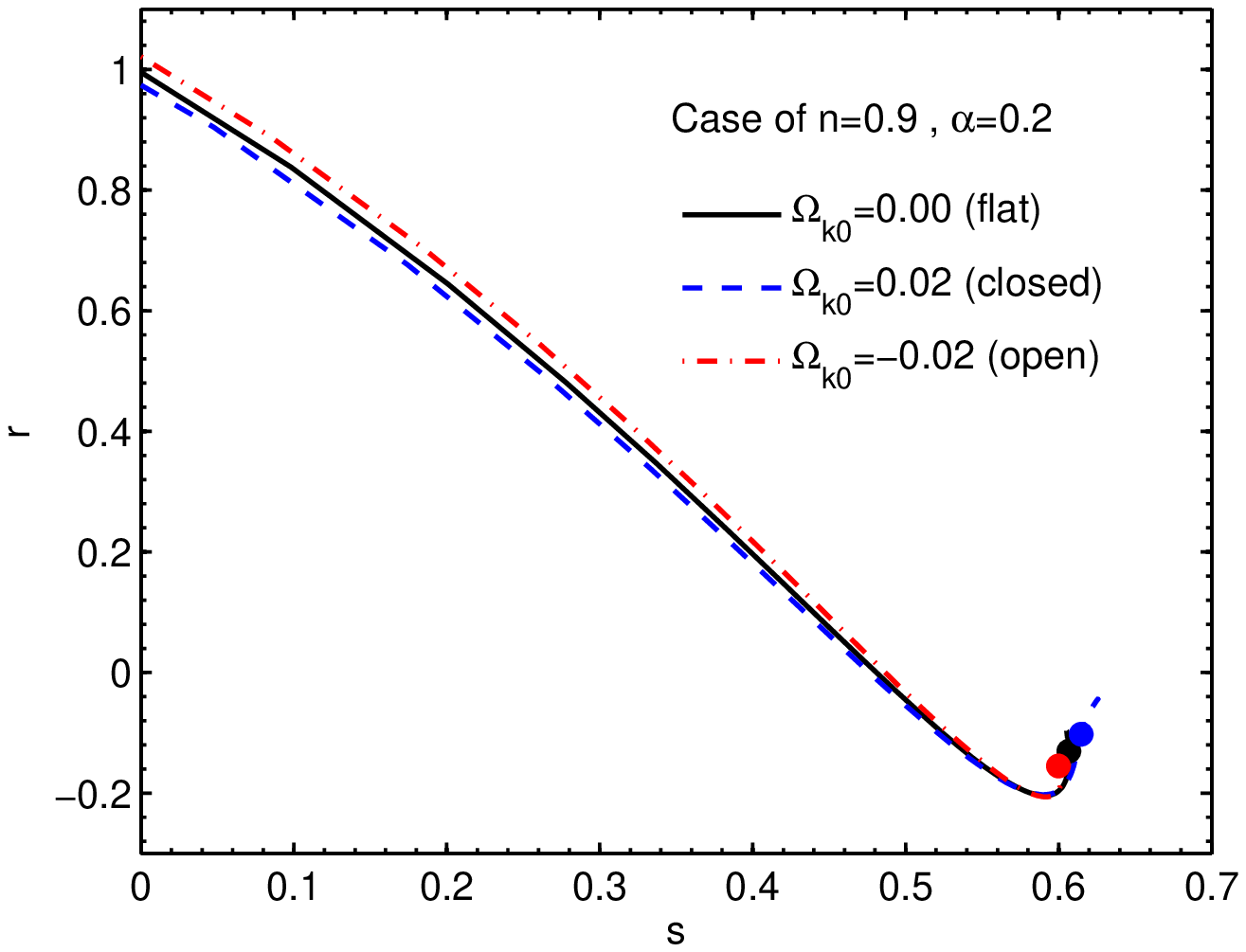} \includegraphics[width=5cm]{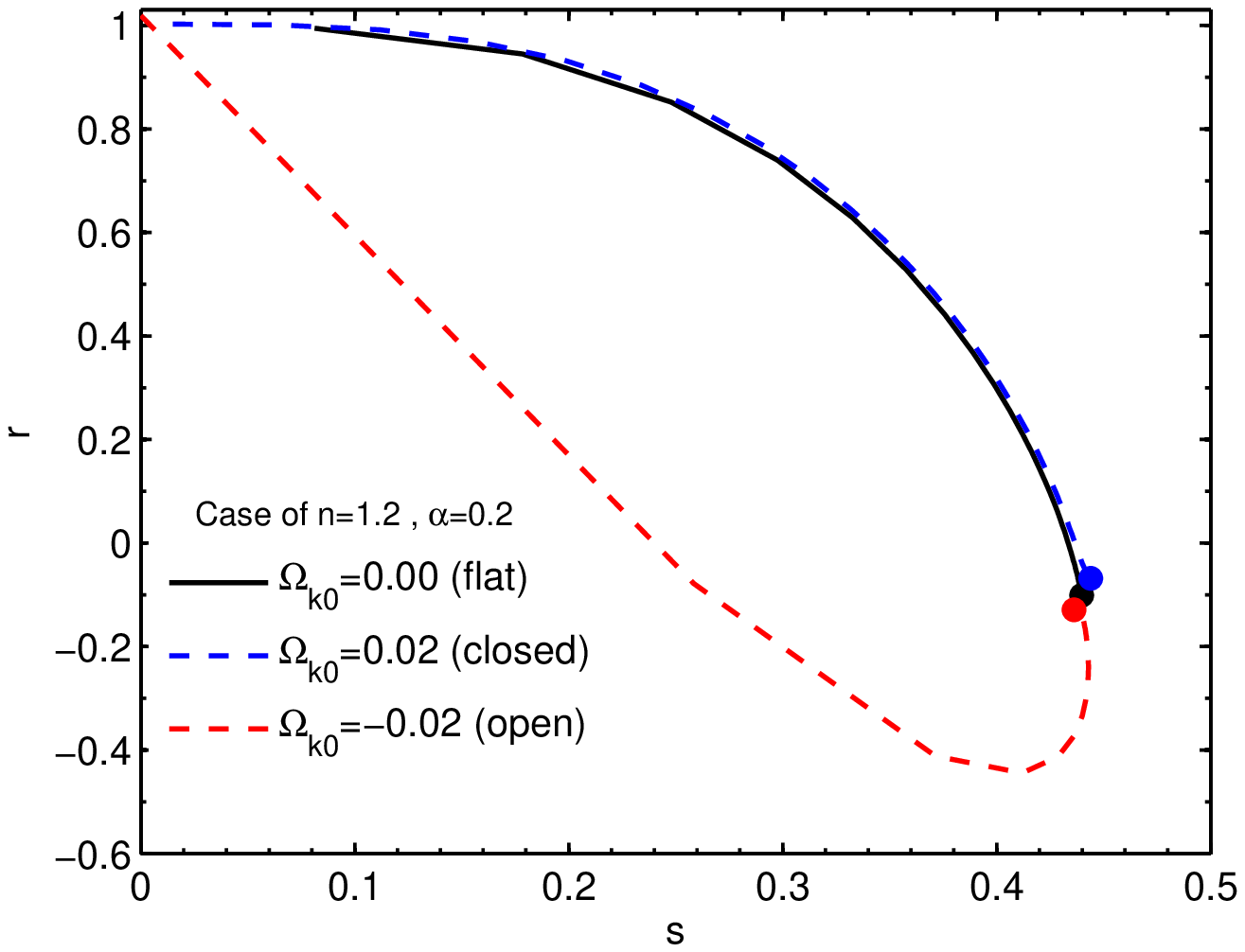}%
\includegraphics[width=5cm]{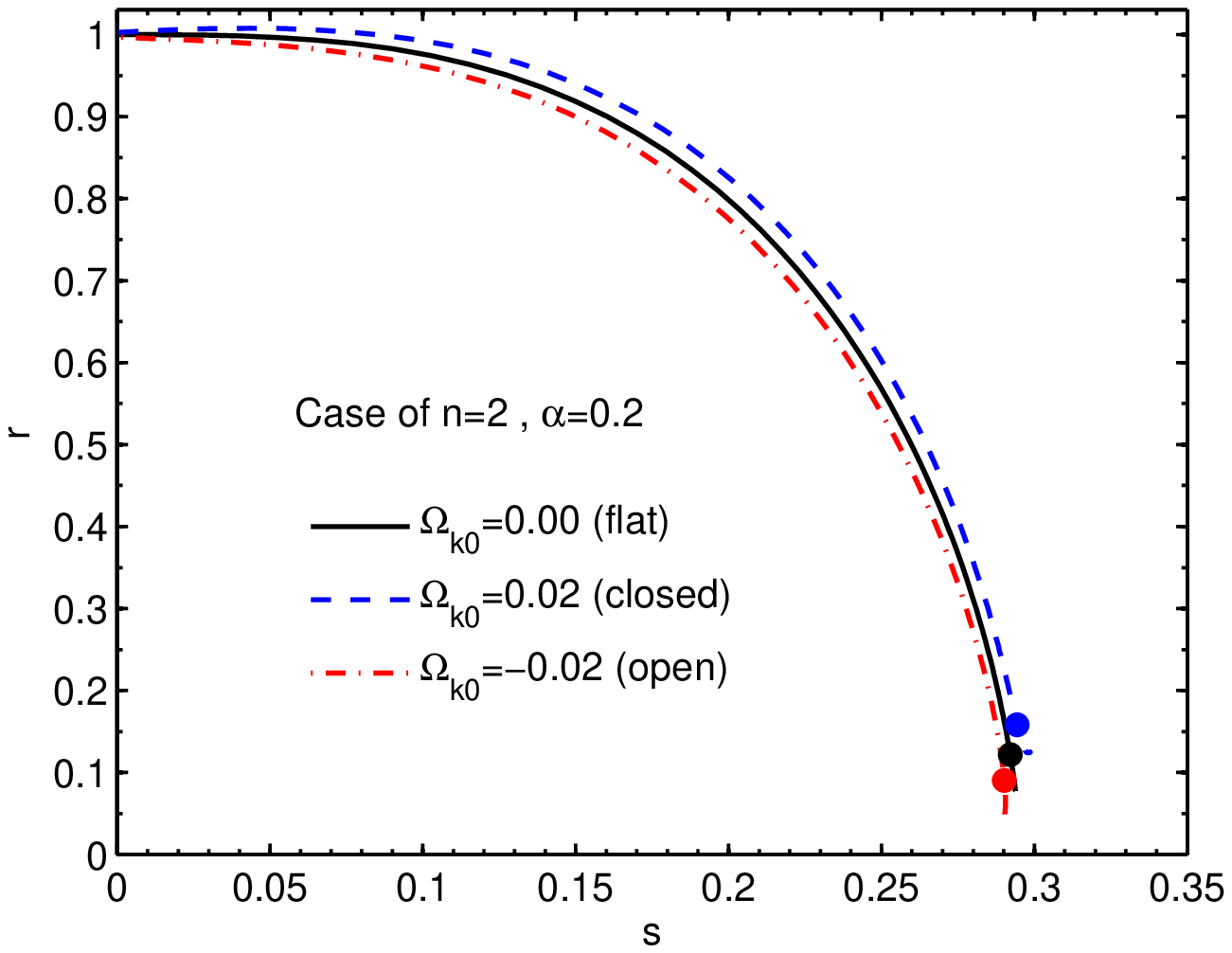}
\caption{The evolutionary trajectories of the statrefinder in the $r-s$ plane for
different model parameters $n$ and $\alpha$ as well as different contribution of spatial curvatures.
In the first arrow panels the ADE model is considered without the
interaction between dark matter and dark energy, while in the second and third arrows the interaction term is taking into account.
The statefinder parameters for the $\Lambda$CAM model corresponds to the fixed point \{r=1,s=0\} in the $r-s$ plane.\\[0pt]}
\end{figure}
\end{center}
 \newpage

\begin{center}
\begin{figure}[!htb]
\includegraphics[width=5cm]{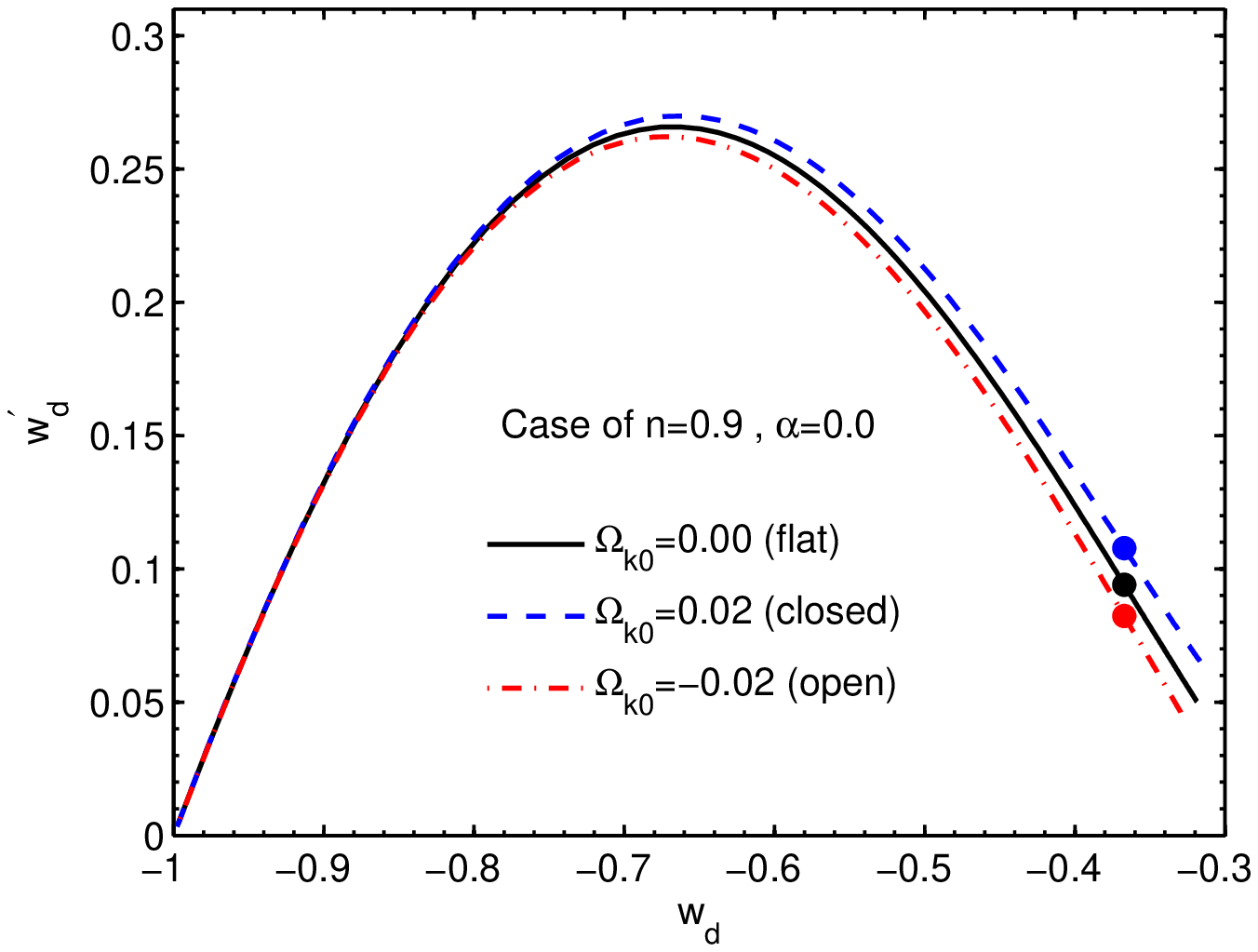} \includegraphics[width=5cm]{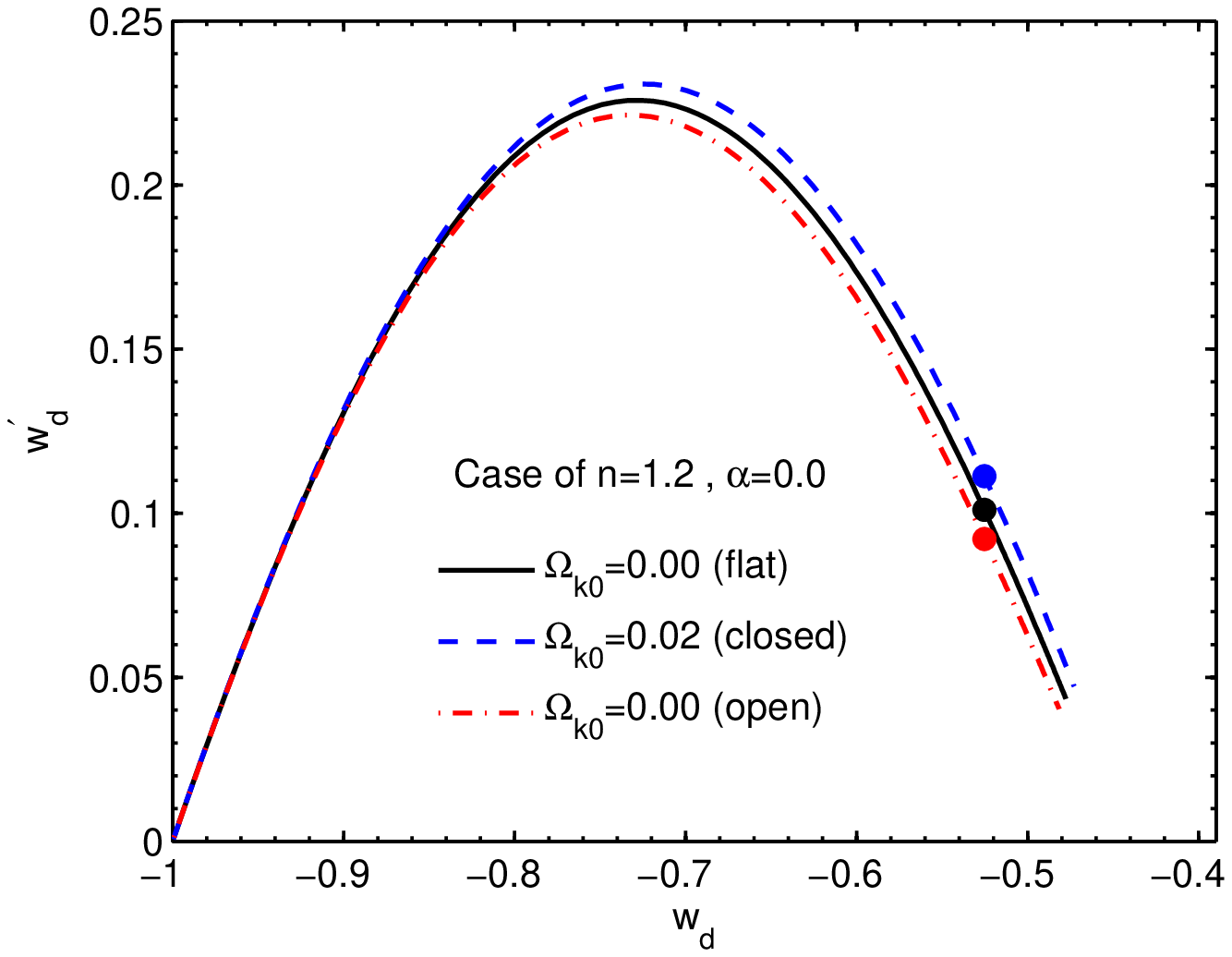} %
\includegraphics[width=5cm]{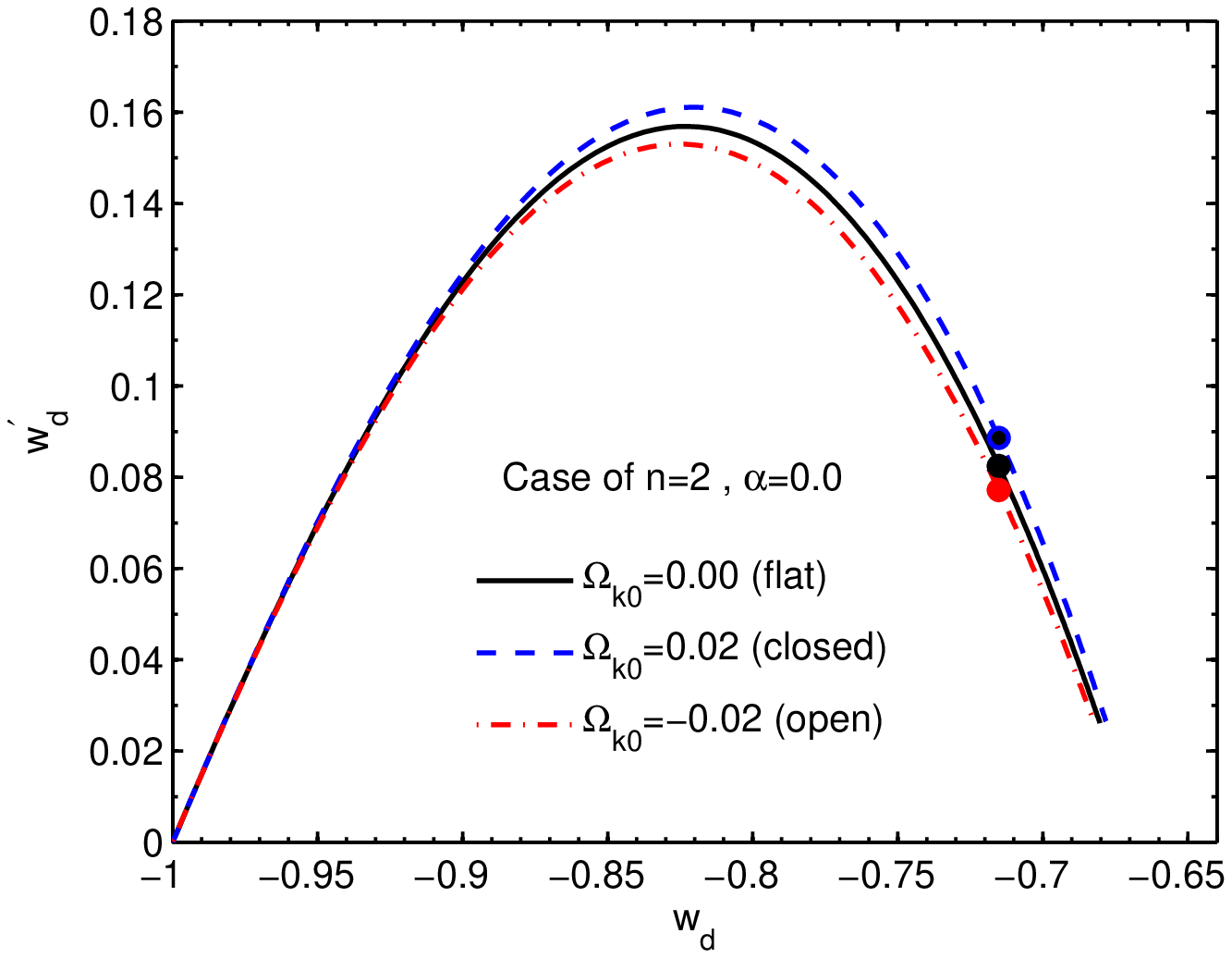} \includegraphics[width=5cm]{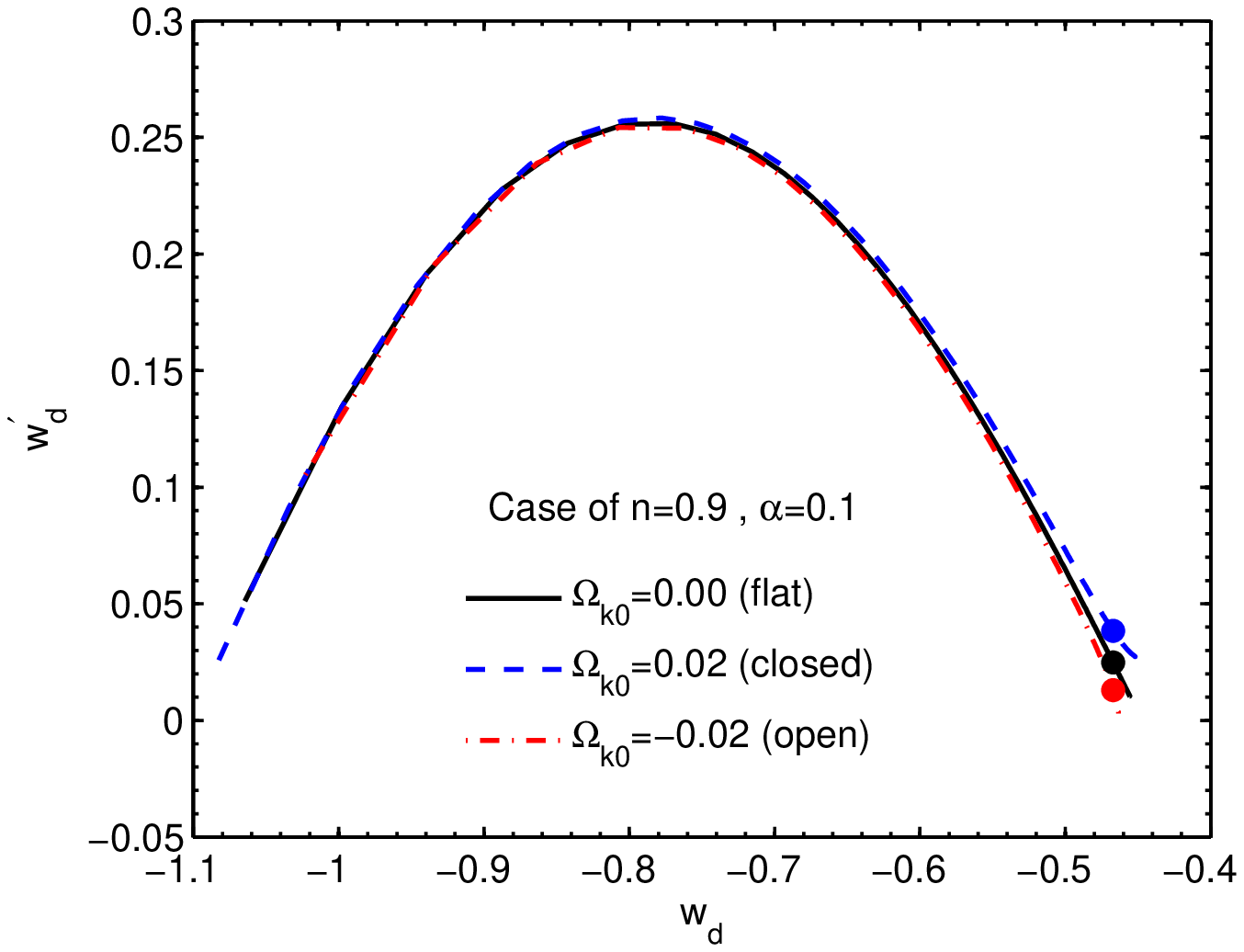} %
\includegraphics[width=5cm]{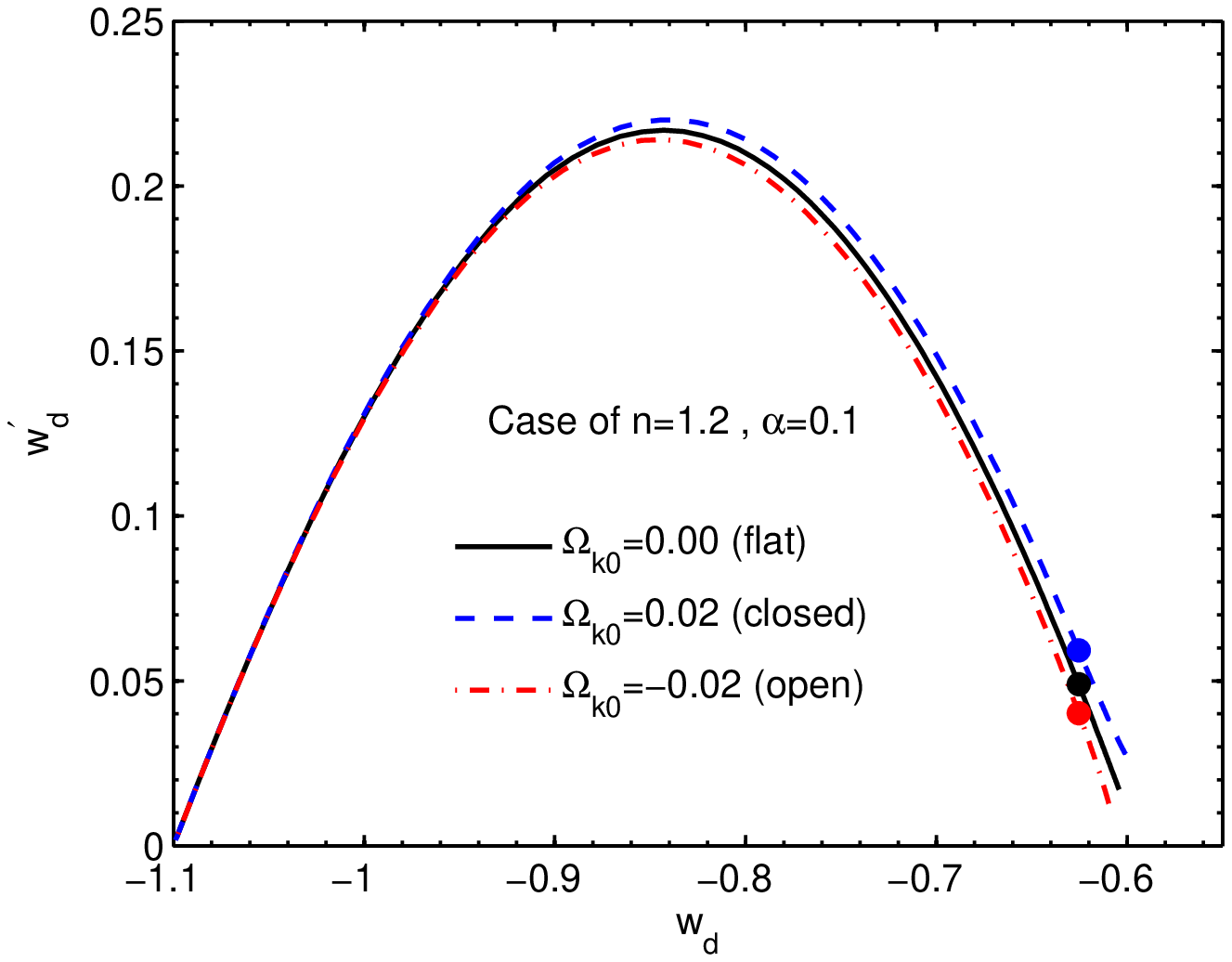} \includegraphics[width=5cm]{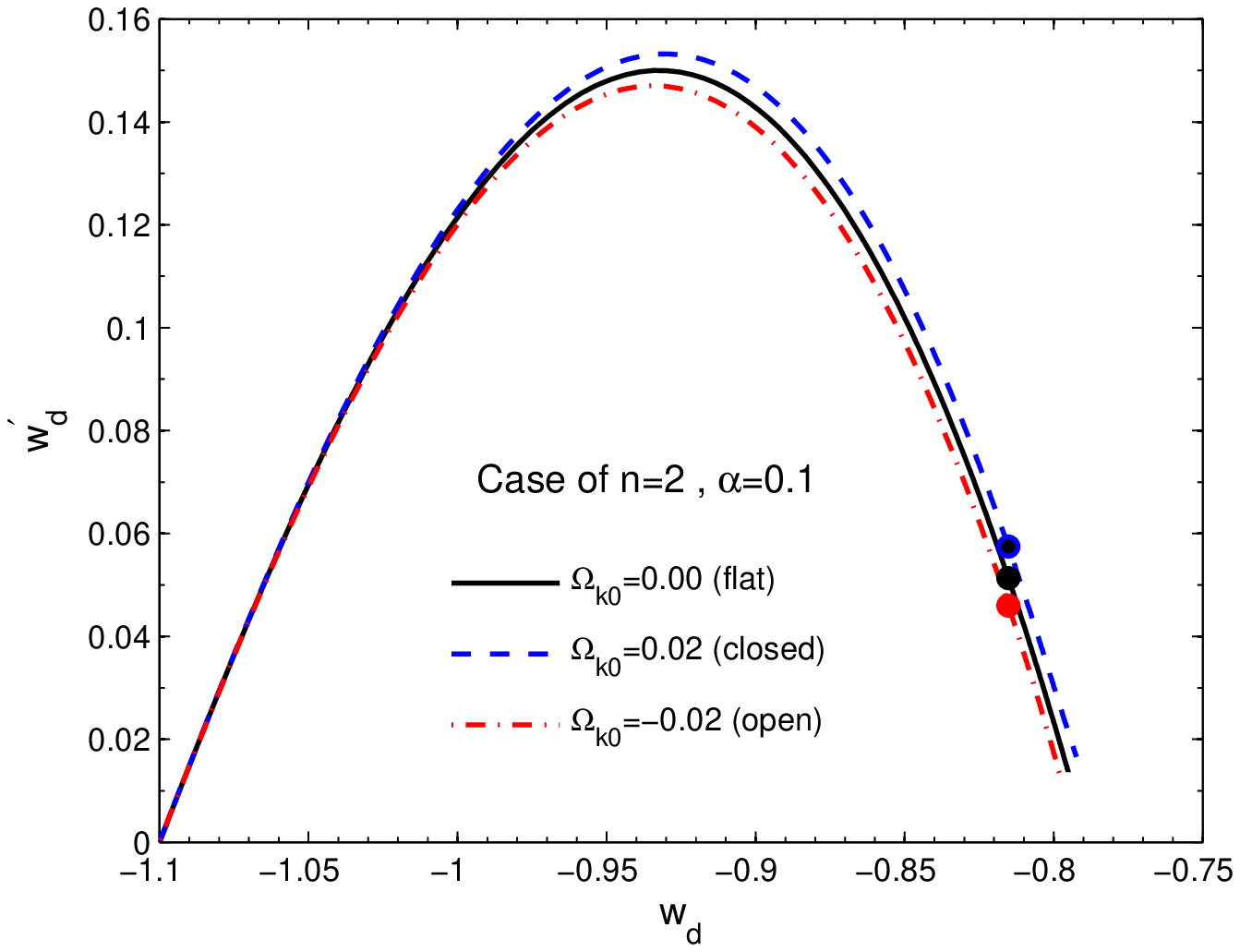} %
\includegraphics[width=5cm]{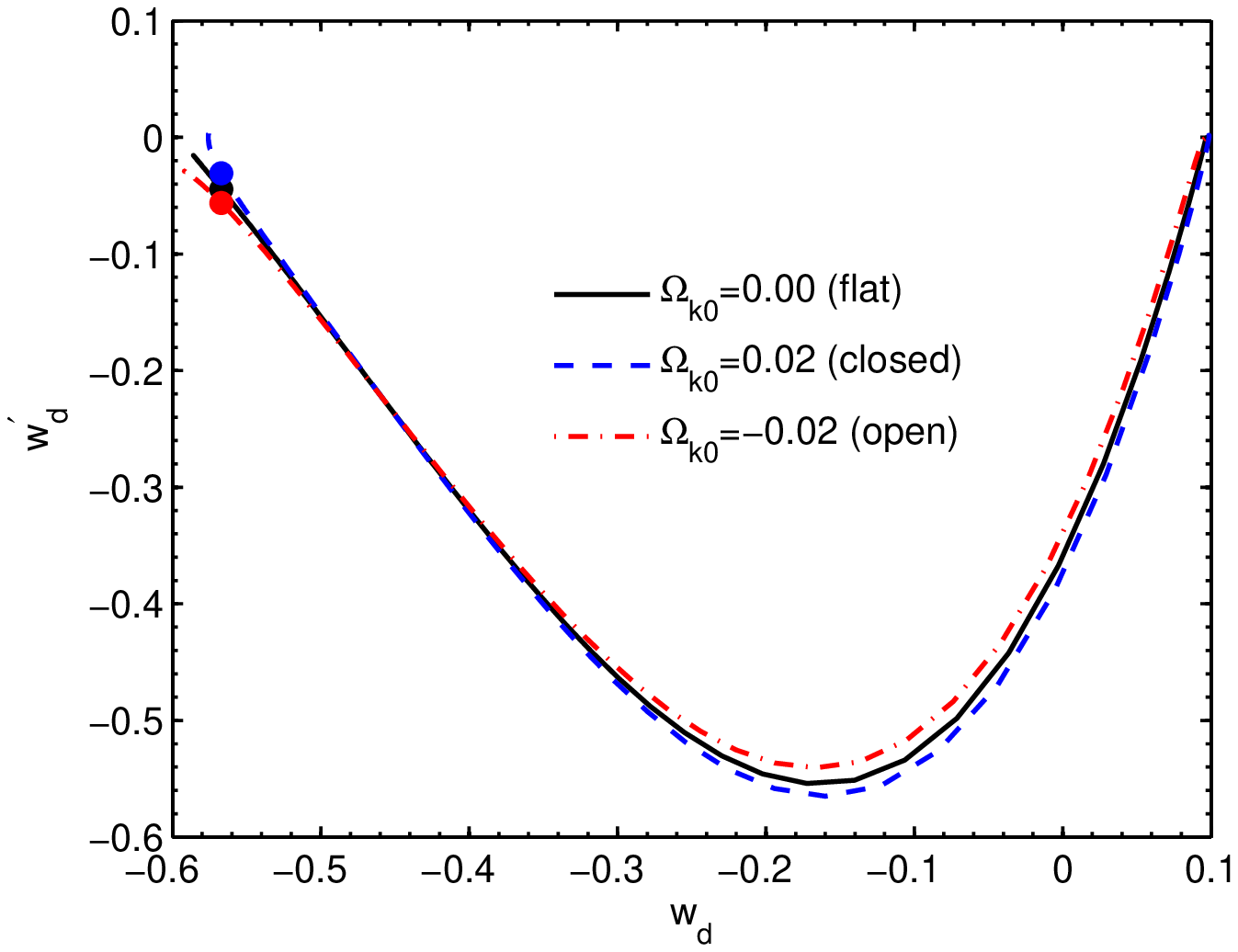} \includegraphics[width=5cm]{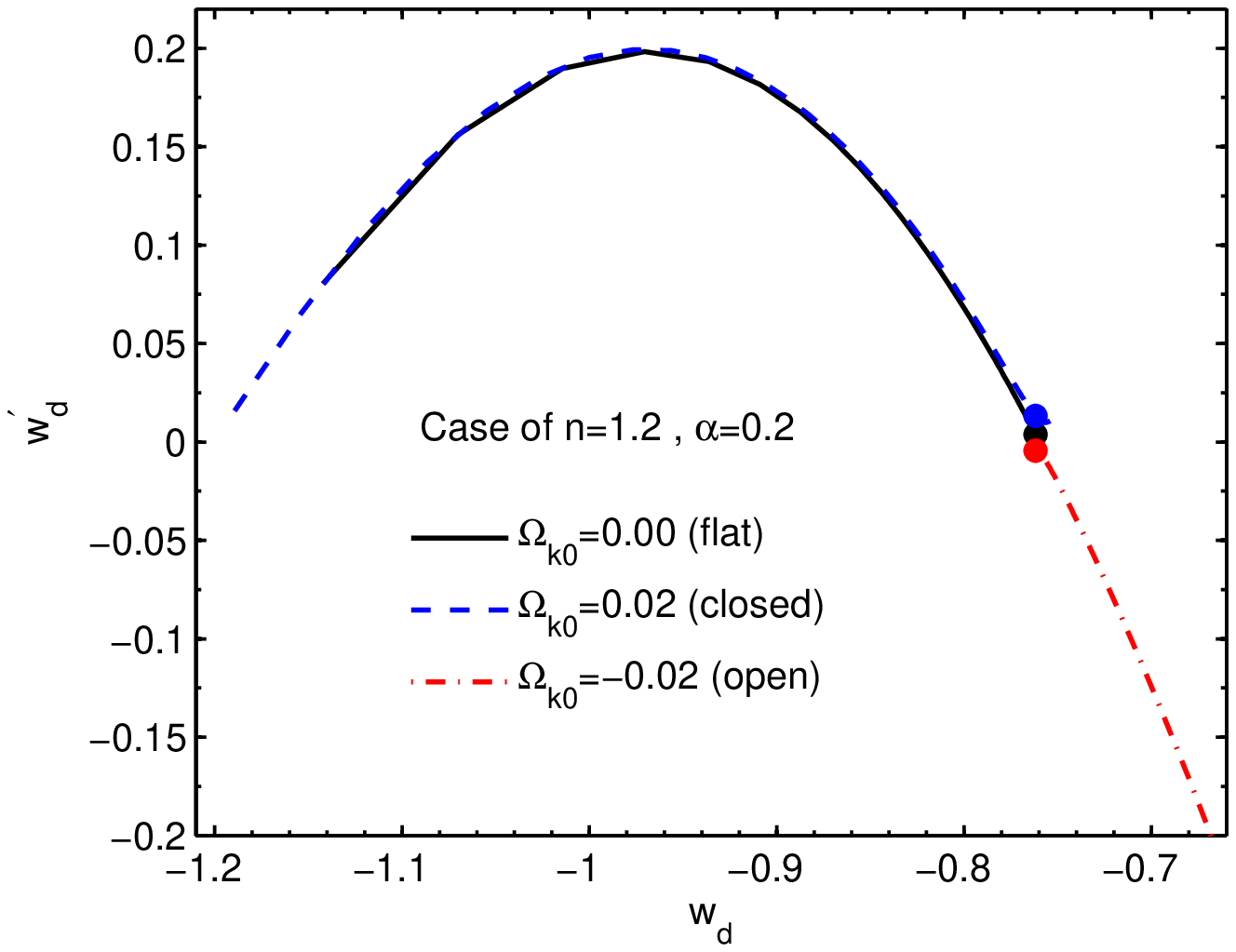}%
\includegraphics[width=5cm]{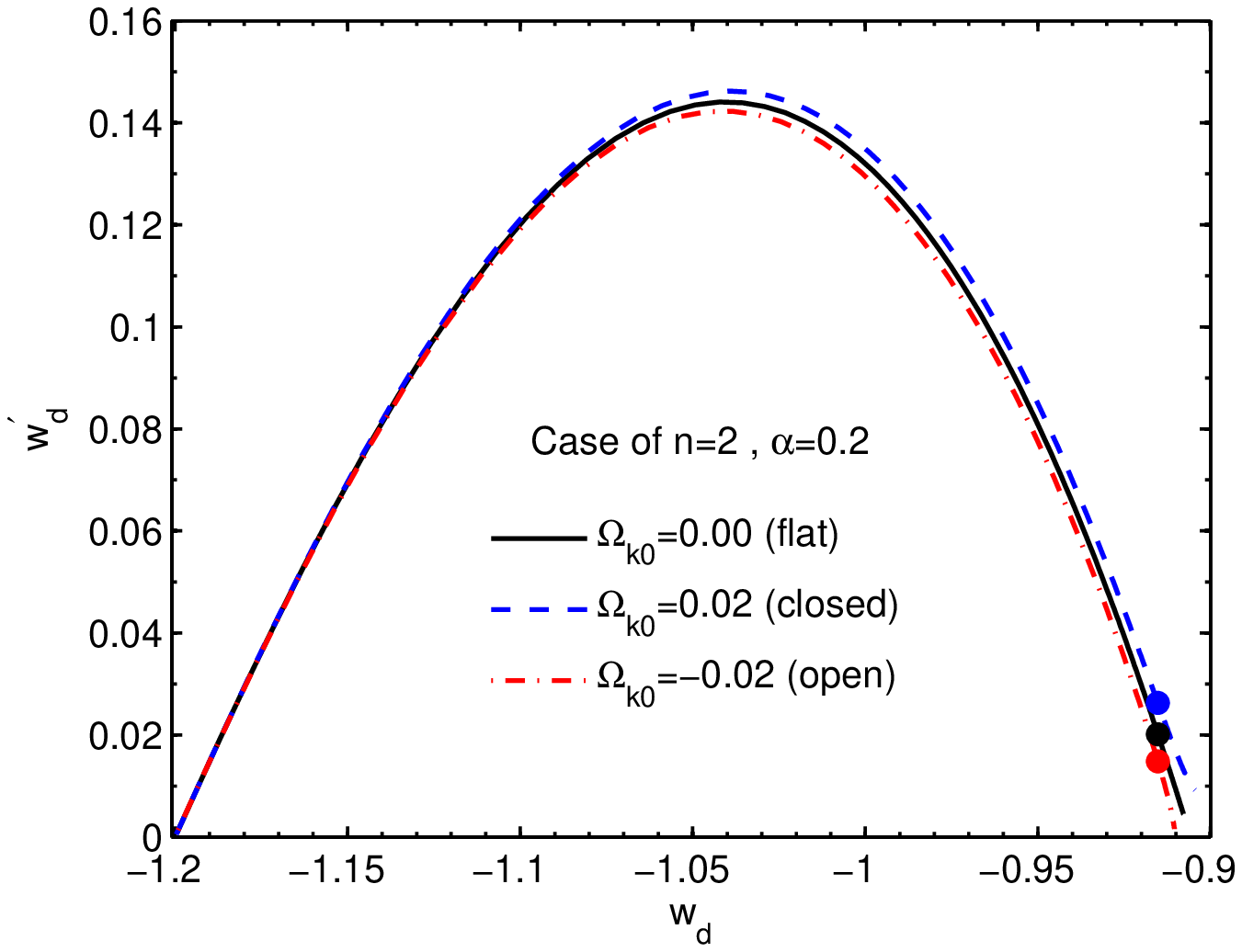}
\caption{The evolutionary trajectories of $\{w_d,w_d^{\prime}\}$ for
different model parameters $n$ and $\alpha$ as well as different
contribution of spatial curvatures. In the first arrow panels the
ADE model is considered without the interaction between dark matter
and dark energy, while in the second and third arrows the
interaction term is taking into account.
The $\{w,w^{\prime}$\} for the $\Lambda$CAM model corresponds to the fixed point $\{w=-1,w^{\prime}=0$\} in the $w-w^{\prime}$ plane.\\[0pt]}
\end{figure}
\end{center}

\end{document}